\documentclass{aa}
\usepackage{graphicx}
\usepackage{txfonts}
\usepackage{natbib}

\bibpunct{(}{)}{;}{a}{}{,}

\def\deg{$^\circ$}

\def\arcsec{$^{\prime\prime}$}

\def\deg{$^{\circ}$}

\def\ltsima{$\; \buildrel < \over \sim \;$}
\def\simlt{\lower.5ex\hbox{\ltsima}} 
\def\gtsima{$\; \buildrel > \over \sim \;$}
\def\simgt{\lower.5ex\hbox{\gtsima}} 

\begin{document}

\idline{A\&A 1, 1--10 (2006)}{1} \doi{}

\title{A kinematic study of the Taurus-Auriga T association}
\author{Claude Bertout\inst{1}
\and Fran\c coise Genova\inst{2}}

\institute{Institut d'Astrophysique, 98bis, Bd. Arago, 75014 Paris,
France \and Centre de Donn\'ees astronomiques de Strasbourg (CDS),
Observatoire de Strasbourg, 11, rue de l'Universit\' e, 67000
Strasbourg, France}

\date{Received June 16, 2006 / Accepted October 12, 2006}
\offprints{Claude Bertout, \email{firstname. lastname@obspm.fr}}

 \abstract
{}
{This is the first paper in a series dedicated to investigating the
kinematic properties of nearby associations of young stellar
objects. Here we study the Taurus-Auriga association, with the
primary objective of deriving kinematic parallaxes for individual
members of this low-mass star-forming region.}
{We took advantage of a recently published catalog of proper motions
for pre-main sequence stars, which we supplemented with radial
velocities from various sources found in the CDS databases. We
searched for stars of the Taurus-Auriga region that share the same
space velocity, using a modified convergent point method that we
tested with extensive Monte Carlo simulations.}
{Among the sample of 217 Taurus-Auriga stars with known proper
motions, we identify 94 pre-main sequence stars that are probable
members of the same moving group and several additional candidates
whose pre-main sequence evolutionary status needs to be confirmed.
We derive individual parallaxes for the 67 moving group members with
known radial velocities and give tentative parallaxes for other
members based on the average spatial velocity of the group. The
Hertzsprung-Russell diagram for the moving group members and a
discussion of their masses and ages are presented in a companion
paper.}
{} 

\authorrunning {C. Bertout \& F. Genova}
\titlerunning {Kinematics of the Taurus-Auriga T~association}
\maketitle

\section{Introduction} \label{IntroductionSection}

To accurately determine the two main physical parameters of nearby
young stellar objects (YSOs), their age and mass, we must know how
far away they are. While determination of distances has been at the
heart of astronomical research for many centuries, we have made
surprisingly little recent progress in this respect, at least for
low-luminosity objects such as the young solar-type T~Tauri stars
(TTSs). This contrasts with more luminous nearby stars, the targets
of the very successful Hipparcos mission, for which accurate
parallaxes are now available. Although Hipparcos observed a few
low-luminosity pre-main sequence (PMS) stars in nearby star-forming
regions \citep[cf.][]{1997A&A...323L..49P}, they were clearly a
challenge for its small telescope. The situation will improve
dramatically with the flight of the Gaia mission, as this satellite
will measure the parallaxes and proper motions of millions of faint
stars. However, Gaia's expected launch date is 2012, so it would be
useful to make some progress in determining the distances of YSOs in
the meantime in order, for example, to better constrain the lifetime
of their disks and the timescales of planet formation, two very
timely research areas for which a precise determination of YSO ages
is urgently required.

Where do we stand today as far as YSO parallaxes are concerned? The
post-Hipparcos situation was discussed by
\cite{1999A&A...352..574B}, who provided new astrometric solutions
of the Hipparcos data for groups of TTSs in various star-forming
regions, thus finding average distances to some YSO groups. The new
distances generally agreed well with previous estimates of the
associated molecular cloud distances based, e.g., on the photometry
of a few bright stars enshrouded in reflection nebulosity. For
distance determinations of the Taurus star-forming region based on
this method, see \cite{1968AJ.....73..233R} and
\cite{1978ApJ...224..857E}.

Although average distances provide valuable information, what we
really need for constraining ages and masses by comparing observed
stellar properties with evolutionary models are the distances to
individual stars of the YSO associations. This is what we are
attempting to do in this work, which focuses on the Taurus-Auriga
T~association. To do so, we use the proper motions of individual
stars.

Several proper motion surveys of the Taurus-Auriga star-forming
region have been performed in the past, e.g., by
\cite{1979AJ.....84.1872J} and \cite{1991AJ....101.1050H}. More
recently, \cite{2005A&A...438..769D} published an all-sky catalog of
proper motions for 1250 YSOs that provides a coherent database for
kinematic studies such as the one we are now embarking on. The
proper motion database for Taurus-Auriga is briefly discussed in
Sect. \ref{CatalogDescriptionSection}.

The procedure that we use here to derive individual parallaxes is as
follows. First, we identify those stars among the Taurus-Auriga
confirmed or suspected YSOs that have the same spatial velocity. In
other words, we look for the group of stars that \emph{defines} the
Taurus-Auriga T association through its common motion in the sky.
While young associations are not expected to be gravitationally
bound, it is well known that their members all share the same motion
in space for several million years before the association dissolves
and loses its identity most notably due to the tidal interactions
caused by Galactic rotation and encounters with other stellar groups
and interstellar clouds \citep[see, e.g., the review
by][]{2002ASPC..285..150B}.

To find the likely association members, we developed our own variant
of the classic convergent point method. This is described in
Sect.~\ref{CPMethodSection}, while Sect.~\ref{MCSims} presents
Monte-Carlo simulations of the moving group search that proved
helpful for choosing computational parameters. We then provide lists
of kinematic members in Sect.~\ref{KinematicAnalysisSection}, where
we also make use of the radial velocity information, when available,
to infer parallaxes and associated error bars for individual stars.
Stellar radial velocities are known only for a limited sub-sample of
association members, but we can compute the average common spatial
velocity of this sub-group, which in turn provides a second way of
determining kinematic parallaxes for all members of the group by
assuming that all stars have the same spatial velocity. These
results are discussed in Sect.~\ref{DiscussionSection}.

At that point, we should be armed with the information needed to
perform a new age and mass determination for members of the moving
group in the T~association. This analysis and its astrophysical
consequences will be presented in a companion paper (Bertout \&
Siess, in preparation).

\section{The sample of Taurus-Auriga stellar objects}
\label{CatalogDescriptionSection}

\begin{figure}
\resizebox{\hsize}{!}{\includegraphics[angle=0]{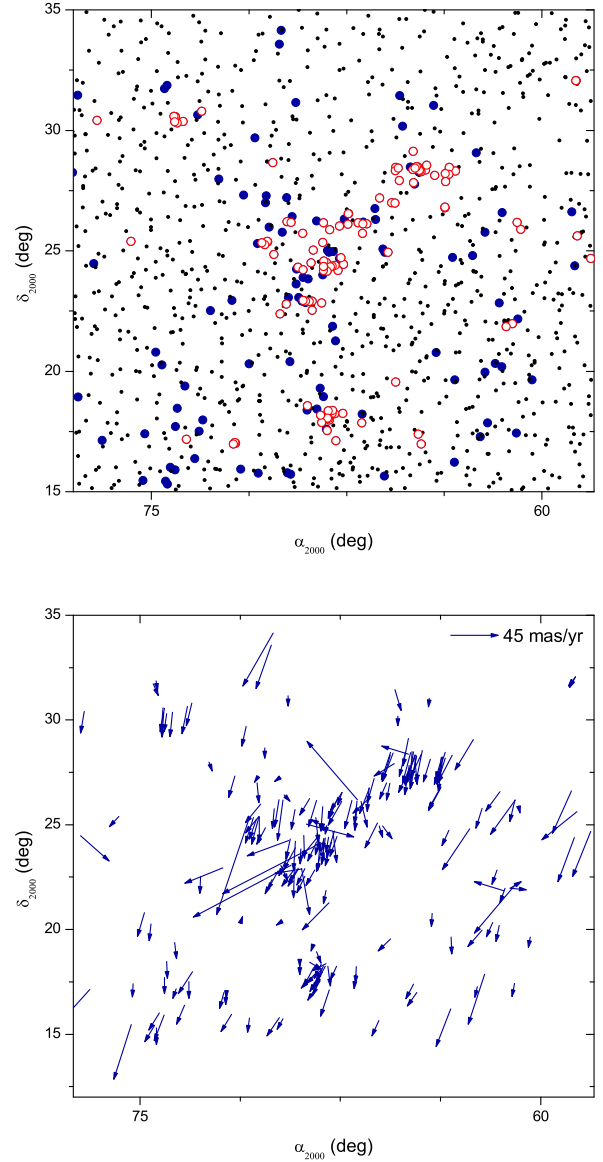}}
\caption[]{Upper panel: Open and filled circles indicate the
positions of Taurus-Auriga stars whose proper motions are given in
the \cite{2005A&A...438..769D} catalog. The open circles denote the
stars that are also included in the \cite{1988cels.book.....H}
catalog of Orion population objects. Dots indicate the locations of
stars included in the Hipparcos catalog. Lower panel: proper motion
vectors of the Taurus-Auriga stars as given in the
\cite{2005A&A...438..769D} catalog.} \label{Fig1}
\end{figure}

The \cite{2005A&A...438..769D} catalog of proper motions for PMS
stars contains 217 stars that are in the general area of the
Taurus-Auriga star-forming region (which roughly spans the range of
coordinates $3^h 50^m$ \simlt $\alpha(2000)$ \simlt $5^h 10^m$ and
$15$\deg \simlt $\delta(2000)$ \simlt $35$\deg).

\subsection{Proper motions of Taurus PMS stars}

\begin{figure}
\resizebox{\hsize}{!}{\includegraphics[angle=0]{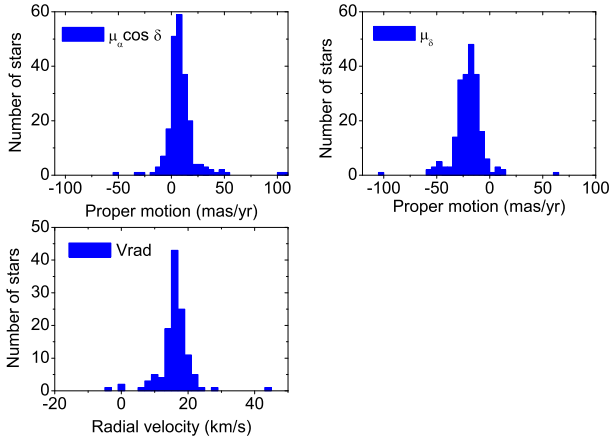}}
\caption[]{Proper motion and radial velocity histograms for the
Taurus-Auriga stars of the \cite{2005A&A...438..769D} catalog.}
\label{Fig2}
\end{figure}

\begin{figure}
\resizebox{\hsize}{!}{\includegraphics[angle=0]{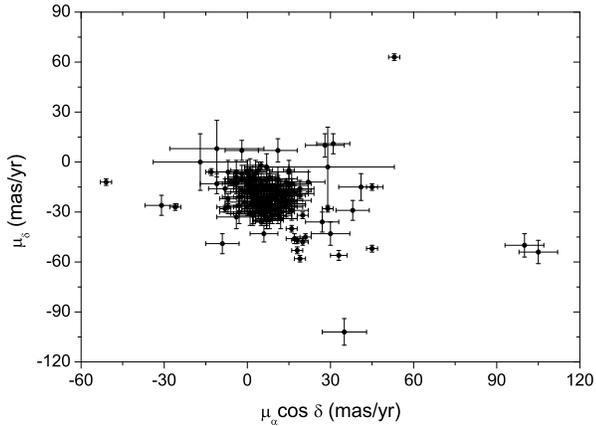}}
\caption[]{Proper motions and associated errors for the 217 stars
considered in this investigation.} \label{Fig3}
\end{figure}

\begin{figure}
\resizebox{\hsize}{!}{\includegraphics[angle=0]{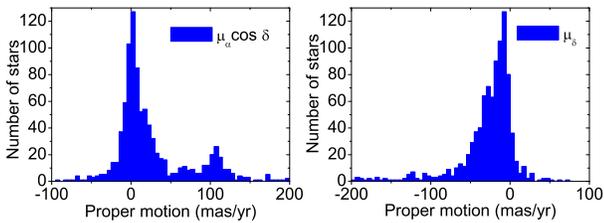}}
\caption[]{Proper motion histograms of the the Hipparcos stars shown
in Fig~\ref{Fig1}.} \label{Fig4}
\end{figure}

\begin{figure}
\resizebox{\hsize}{!}{\includegraphics[angle=0]{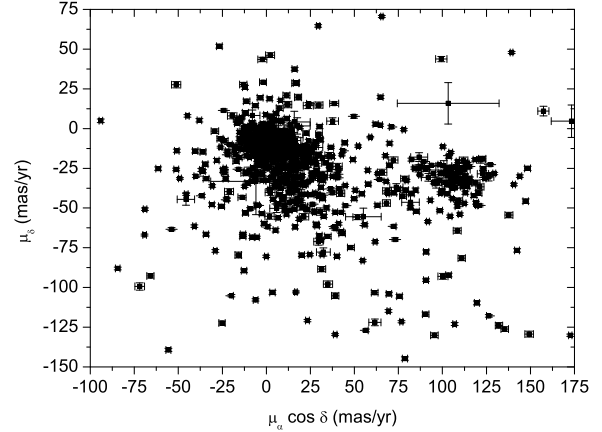}}
\caption[]{Proper motions and associated errors for the Hipparcos
stars shown in Fig~\ref{Fig1}.} \label{Fig5}
\end{figure}

The upper panel of Fig.~\ref{Fig1} displays the location of
Taurus-Auriga stars contained in the \cite{2005A&A...438..769D}
catalog. One recognizes the familiar grouping of YSOs in the
vicinity of Taurus molecular cores: L1551 at approximately $\alpha
(2000) = 4^h 32^m$ and $\delta (2000) = 18$\deg. North of L1551 one
finds the L1536 and L1529 groups, while in the northwest of L1529
one has the L1495 group and in the northeast the HCL2 group. The few
Auriga stars present in the sample are around $\alpha (2000) = 5^h$
and $\delta (2000) = 30$\deg. Individual objects at the periphery of
the molecular cloud are mainly weak emission-line TTSs discovered
through their X-ray emission.

For comparison, we also show in Fig.~\ref{Fig1} the stars observed
by the Hipparcos satellite in the same region. Because of
extinction, the density of Hipparcos stars strongly decreases in the
vicinity of the Taurus-Auriga molecular clouds where our targets are
located. As we discuss later, only a few PMS stars in the region are
common to both the Hipparcos \citep{1997A&A...323L..49P} catalog and
the \cite{2005A&A...438..769D} catalog of presumably young stars.

The lower panel of Fig. \ref{Fig1} shows the proper motion vectors
for all objects in the \cite{2005A&A...438..769D} catalog. Although
a clear convergent point for these proper motions is not readily
apparent, one notices that many proper motion vectors seem to point
toward the lower left-hand corner of the figure.

Figure~\ref{Fig2} presents histograms of the proper motion
components $\mu_\alpha \cos \delta$ and $\mu_\delta$, as well as a
histogram of the radial velocities for those 127 objects for which
we could find a published value (see below). The average values and
standard deviations for the proper motions of the full sample are

\medskip
\noindent $\left\{ \begin{array}{lll}
\mu_\alpha \cos \delta & = & 8.20 \pm 14.45 \: \rm{mas/yr} \\
\mu_\delta & = & -20.82 \pm 13.84 \: \rm{mas/yr} \\
\end{array}\right.$
\medskip

If we then consider the sub-sample of 127 stars with known radial
velocities, we have

\medskip
\noindent $\left\{ \begin{array}{lll}
\mu_\alpha \cos \delta & = & 7.16 \pm 8.55 \: \rm{mas/yr} \\
\mu_\delta & = & -20.91 \pm 10.31 \: \rm{mas/yr} \\
v_{rad} & = & 16.03 \pm 6.43 \: \rm{km/s},
\end{array}\right.$
\medskip

\noindent where we note a lower dispersion of the proper motion
measurements, due in particular to the fact that radial velocities
have been measured primarily in the brightest and most confirmed
members of the PMS population. We come back to this point in
Sect.~\ref{KinematicAnalysisSection}.

Figure~\ref{Fig3} displays all proper motion values and their
associated uncertainties. Proper motion values cluster around the
average values, albeit with considerable scatter, while a few stars
have highly discrepant proper motions, often with large error bars.
As discussed in Sect.~\ref{KinematicAnalysisSection}, some of the
stars included in the \cite{2005A&A...438..769D} catalog are likely
to be field stars, which explains the discrepant values. However, a
majority of stars with measured radial velocities are confirmed PMS
stars.

\subsection{Caveats}

One difficulty with the investigation envisioned here is the
apparent similarity between the proper motions of field stars, as
represented by the Hipparcos\footnote{Note that we assume implicitly
throughout this section that Hipparcos stars are reasonable proxies
for the fainter field star population that could contaminate the PMS
catalog.} targets of Fig.~\ref{Fig1}, and the proper motions of our
target stars.  To illustrate this, we show histograms of the two
proper motion components for those Hipparcos stars in
Fig.~\ref{Fig4}. We also depict the proper motion data for these
objects, together with their error bars,  in Fig.~\ref{Fig5}.

Although the similarities with Figs.~\ref{Fig2} and \ref{Fig4} are
obvious, there are some differences between the proper motion
distributions that become more evident when one studies the
histogram shapes in more detail. We fitted these data with Gaussian
curves and while it was easy to fit the proper motions of the
\cite{2005A&A...438..769D} objects with a single Gaussian, three
different components were needed to fit the Hipparcos data in a
satisfactory way. Actually, the three components are easily seen in
Fig.~\ref{Fig5}, and Table~\ref{PMfits} summarizes the properties of
the various Gaussian curves that best fitted the data, as well as
the squared coefficient of multiple correlation $R^2$ and the
reduced $\chi^2$ values for the overall fits. The
\cite{2005A&A...438..769D} proper motions are marginally compatible
with the proper motion values of Peak \#1 of the Hipparcos proper
motions but with much smaller dispersions, and the overall proper
motions distributions of the two samples appear quite different.
Peaks \#2 and \#3 correspond approximately to the proper motions of
the Pleiades and Hyades clusters. We see from Fig.~\ref{Fig1} that
several stars in the \cite{2005A&A...438..769D} catalog appear to be
probable members of the Pleiades, while two stars have proper
motions consistent with Hyades membership. It is therefore clear
that the catalog is contaminated by non pre-main sequence stars to
some extent (cf. Sect.~\ref{KinematicAnalysisSection}).

To further quantify the differences between the proper motion
distributions, we performed a Kolmogorov-Smirnov test on the two
components of the proper motion for the \cite{2005A&A...438..769D}
and Hipparcos samples of stars. We found that the probability of the
distributions of $\mu_\alpha \cos \delta$ components for both
samples being drawn randomly from the same parent distribution is $8
\cdot 10^{-14}$, while it is $4 \cdot 10^{-7}$ for the $\mu_\delta$
components.

\begin{table}
\caption{Gaussian fits to the proper motion components of stars in
the Taurus-Auriga region. The data were binned as in
Figs.~\ref{Fig2} and \ref{Fig4}.}\label{PMfits}
\begin{center}
\tiny{
\begin{tabular}{lccccc}
\hline\hline PM comp.  & Peak \#  &  PM & FWHM  & $R^2$ & $\chi^2$   \\
 & & mas/yr & mas/yr & & \\
\hline
\multicolumn{6}{c}{Ducourant et al. sample} \\
\hline $\mu_\alpha \cos \delta $& - & 6.93 $\pm$ 0.14 & 12.11 $\pm$
0.28 &
0.99 &  0.77  \\
$\mu_\delta$ & - & -19.29 $\pm$ 0.20 & 14.85 $\pm$ 0.41 & 0.99 & 0.87  \\
\hline
\multicolumn{6}{c}{Hipparcos sample} \\
\hline
 & 1 & 0.93 $\pm$ 0.26 & 10.07 $\pm$ 0.67 \\
$\mu_\alpha \cos \delta$ & 2 &8.89 $\pm$ 0.95 & 39.38 $\pm$ 1.76 & 0.98 &  12.63 \\
  &  3 & 103.74 $\pm$ 2.07 & 36.54 $\pm$ 4.17 \\
 & 1 & -7.77 $\pm$ 0.25 & 10.98 $\pm$ 0.73 \\
$\mu_\delta$ & 2 & -46.66  $\pm$ 11.34 & 102.12 $\pm$ 18.19  & 0.99 &  8.84\\
 & 3 & -22.98 $\pm$ 1.26 & 31.45 $\pm$ 1.98\\

\hline
\end{tabular}
}
\end{center}
\end{table}

\begin{figure*}
\resizebox{\hsize}{!}{\includegraphics[angle=0]{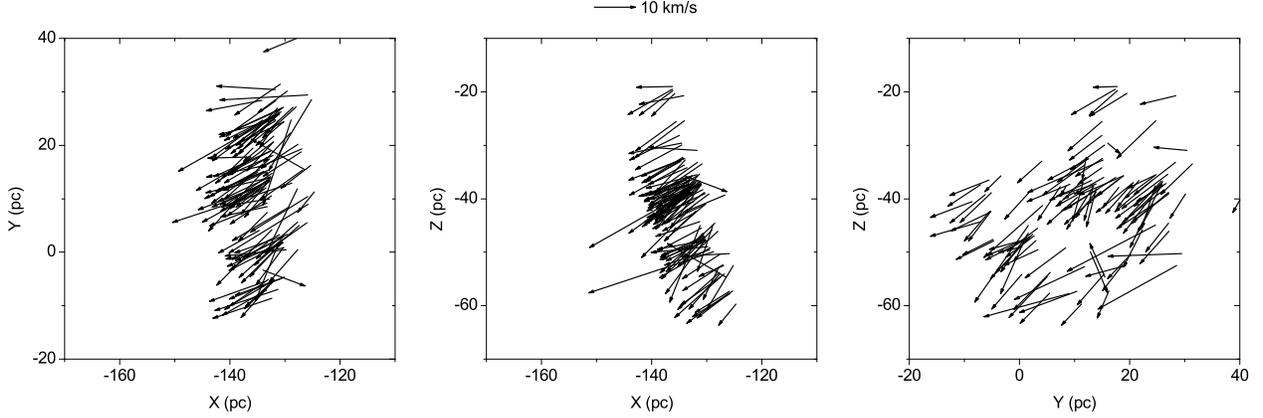}}
\caption[]{Projection on the 3 planes XY, XZ, and YZ of the spatial
velocities of 127 Taurus stars with known radial velocities,
assuming a common distance of 139pc. The velocity vectors originate
at the stars' coordinates in the XYZ reference frame drawn in
Fig.~\ref{Fig7}. As in this figure, the spatial scales were chosen
for ease of comparison with the results of this investigation.}
\label{Fig6}
\end{figure*}

While these differences raise the hope that the following analysis
will allow us to recognize the common motion of PMS objects even in
the presence of field stars, it is also clear from the above that
field stars must exist that share approximately the same proper
motions as the \cite{2005A&A...438..769D} sample. Some of them could
possibly contaminate the catalog, although it is meant to include --
at least in principle -- only PMS stars. Because we cannot
kinematically distinguish between true members of the association
and at least some field stars, it is crucial to scrutinize the
catalog for non-PMS objects and to screen out these possible
interlopers before we look for a moving group of young stars. We
come back to this issue in Sect.~\ref{KinematicAnalysisSection}.

Another worry when dealing with the kinematics of young stars in
Taurus is the high degree of stellar multiplicity in that
star-forming region, where the duplicity reaches about 49\%, a
factor 1.9 larger than for solar-type field stars
\citep{1998A&A...331..977K}. One of the incentives cited by
\cite{2005A&A...438..769D} for preparing a proper motion catalog of
PMS objects is that \emph{``dedicated work }[on objects that have
close companions] \emph{is necessary to derive more reliable proper
motions",} and this is indeed what makes this catalog so valuable.
Many binaries in Taurus are relatively wide pairs, often with large
flux ratios, for which individual proper motion determinations of
the components can be done. There are also a number of binaries with
sub-arcsecond separations \citep [cf. ][]{2001A&A...369..249W} that
remain unresolved by \cite{2005A&A...438..769D} and are therefore
unlikely to significantly affect their results. The effect of
duplicity on the catalog's proper motions is expected to be most
severe for relatively close binaries with a few arcsecond separation
that were barely resolved by \cite{2005A&A...438..769D}. These
systems are identified in the catalog with a mention AB that we kept
in our tables. It indicates that the given proper motion is
representative of the binary's motion.

\subsection{Radial velocities of Taurus PMS stars}

The \cite{1988cels.book.....H} catalog contains all radial velocity
values known prior to 1988. In order to access the more recent
measurements, we searched the CDS databases using some of the data
mining tools available on the CDS site. The search made use of a
prototype implementation of the Unified Content
Descriptors\footnote{UCDs are standardized descriptors of
astronomical quantities defined by the International Virtual
Observatory Alliance; cf. {\tt
http://www.ivoa.net/Documents/latest/UCDlist.html.}} in the VizieR
database.

We found radial velocity information for only 127 stars of our
sample, which is quite surprising given the many investigations of
the Taurus YSOs that can be found in the literature. Besides
\cite{1988cels.book.....H}, the main sources of radial velocity
measurements for Taurus-Auriga stars are \cite{1949ApJ...110..424J},
\cite{1986ApJ...309..275H}, \cite{1987AJ.....93..907H},
\cite{1988AJ.....96..297W}, \cite{1997A&A...325..647N}, and
\cite{2000A&A...359..181W}. While we might have missed some data
that are not included in the CDS databases or are not known to us
otherwise, it is also obvious that investigations of Taurus-Auriga
have often focused on the same relatively bright stars in the
region, and no systematic, high-precision radial velocity survey has
been performed to date in spite of the availability of efficient
spectrographs on medium-sized telescopes.

\subsection{Galactic velocities of the \cite{2005A&A...438..769D} sample}

\begin{figure*}
\sidecaption
\includegraphics[width=12cm]{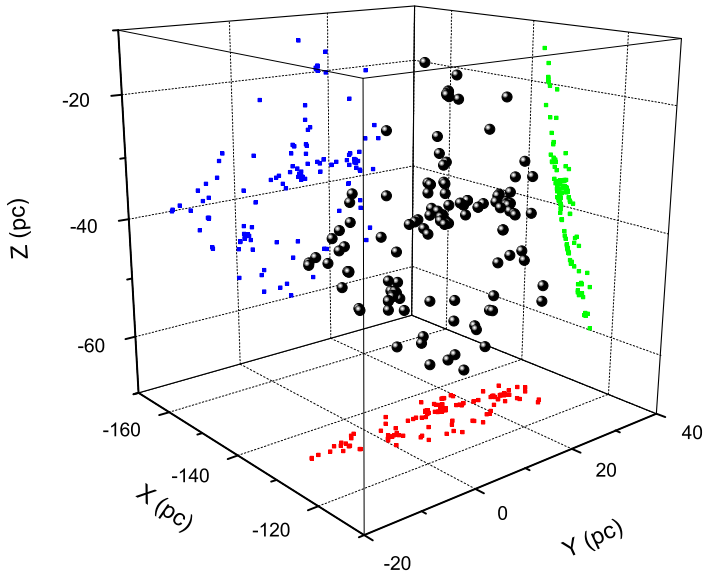}
\caption[]{The 3D spatial distribution of the 127 Taurus stars with
known radial velocities, assuming a common distance of 139 pc, in a
cube with sides equal to 60 pc. The XYZ reference frame is defined
in the text; the Sun is located at its origin. The ball-like symbols
represent individual stars whose projections on the 3 planes are
shown as small squares with different colors.} \label{Fig7}
\end{figure*}

We now compute the Galactic positions and velocities of the 127
stars for which we know radial velocities, using the method
delineated in \cite{1987AJ.....93..864J}. The Galactic positions are
defined on an $XYZ$-grid with the origin at the Sun. There, $X$
points to the Galactic center, $Y$ points in the direction of
Galactic rotation, and $Z$ points to the Galactic North pole. The
projection of the Galactic velocity on the same grid defines the
components $U$, $V$, and $W$.

We plotted the projections of the Galactic velocities onto the three
planes $XY$, $XZ$, and $YZ$  in Fig.~\ref{Fig6}, mainly for the
purpose of comparison with the results obtained in the following
sections. Using the direction of the equatorial coordinate system
south-north axis indicated above, one can check that the directions
of velocity projections in the $YZ$-plane follow the directions of
the proper motion vectors of Fig.~\ref{Fig1}, as expected. Although
there is an obvious scatter in the velocity directions, it appears
from these projections that a number of stars may have similar space
motions, thus providing an incentive for the following
investigation.

Figure~\ref{Fig7} is a 3D cube with sides equal to 60 pc showing the
Galactic positions of the Taurus stars when, following
\cite{1999A&A...352..574B}, we assume an average distance of 139 pc
for the Taurus complex. With this average distance, the stars
populate an $XZ$-plane at $X \approx -132$ pc, and one recognizes
the T~Tauri stellar groupings in this plane. Note that the
south-north axis of the equatorial coordinate system runs
approximately parallel to the direction of the diagonal from
$Z=-70$, $Y=-20$ to $Z=-10$, $Y=40$. The size of the 3D cube was
chosen for comparison with the results of
Sect.~\ref{KinematicAnalysisSection}.

\begin{figure}
\resizebox{\hsize}{!}{\includegraphics[angle=0]{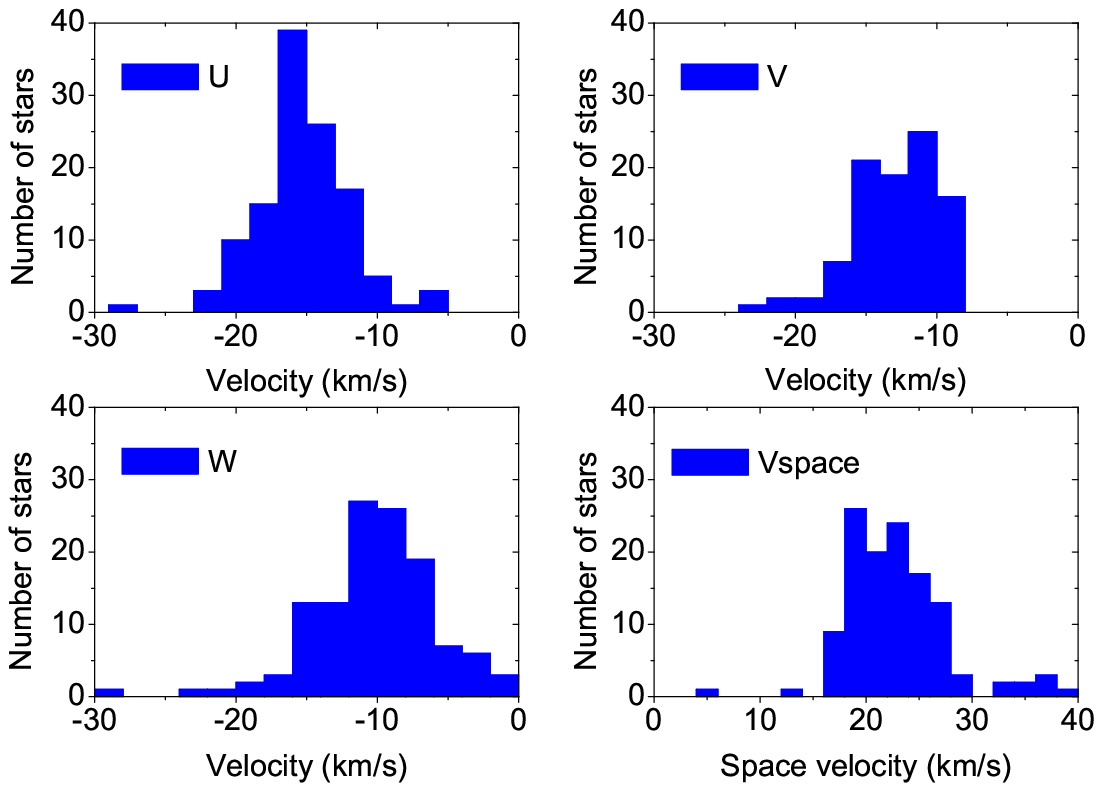}}
\caption[]{Histograms of the U, V, W Galactic velocity components,
as well as the magnitudes of the spatial velocities for 127 Taurus
stars with known radial velocity, assuming a common distance of
139pc.} \label{Fig8}
\end{figure}

Finally, we have plotted histograms of the $U$, $V$, $W$, and
$V_{rad}$ values in Fig.~\ref{Fig8}. Their respective average and
standard deviation values are

\medskip
\noindent $\left\{ \begin{array}{lll}
U & = &  -15.37 \pm 6.00 \: \rm{km/s}  \\
V & = &  -11.74 \pm 6.50 \: \rm{km/s}  \\
W & = &   -9.89 \pm 4.94 \: \rm{km/s}  \\
V_{space} & = & 23.20 \pm  5.98 \: \rm{km/s}.
\end{array}\right.$
\medskip

\section{Method of analysis}\label{CPMethodSection}

We have developed our own variant of the well-known convergent point
method for finding members of moving groups that share the same
space motion. The original method goes back to
\cite{1916cels.book.....C.} and was further developed by several
workers, including \cite{1950ApJ...112..225B},
\cite{1971MNRAS.152..231J}, and more recently by
\cite{1999MNRAS.306..381D}, who adapted it so as to take full
advantage of the Hipparcos data. We review briefly the method before
outlining the variant we developed to deal with the problem at hand.

\subsection{The classic convergent point method}

The convergent point method is based on the fact that the
proper-motion vectors of a group of stars moving with the same
motion in space appear to an observer to converge to a specific
point of the sky plane, called the convergent point (CP hereafter).

The two components of the proper motion vector $\vec{\mu}$ in the
equatorial coordinate system, $\mu_\alpha \cos \delta$ and
$\mu_\delta$, define the tangential velocity vector $\vec{v_{tan}}$
through the relation
\begin{equation}\label{vtan}
\vec{v_{tan}} \equiv \frac{A \: \vec{\mu}}{\pi}
\end{equation}
where $\pi$ is the parallax in milliarcseconds (mas) and where $A =
4.74047$ km yr/s, the ratio of one astronomical unit in km to the
number of seconds in one Julian year, is the constant needed for
adjusting the units on both sides of the equation.

\begin{figure}
\resizebox{\hsize}{!}{\includegraphics[angle=0]{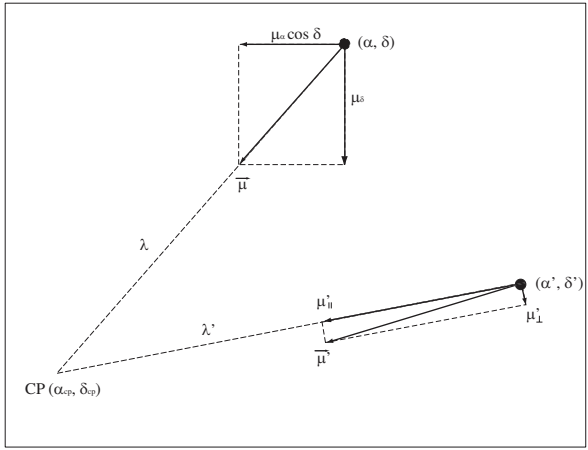}}
\caption[]{Schematic representation of the various proper motion
components used in this analysis. See text for details.}
\label{Fig9}
\end{figure}

If we denote the modulus of the space velocity vector by \(
\left|\;\vec{v}\;\right| \) and, for the time being, neglect
measurement errors and a possible internal cluster velocity
dispersion, we have
\begin{equation}\label{BasicEq}
\left|\;\vec{\mu}\;\right| =  \frac{\pi \left|\;\vec{v}\;\right|
\sin \lambda }{A}
\end{equation}
where $\lambda$ is the angular distance between the position
($\alpha, \delta$) of a given star in the moving group and the
position ($\alpha_{CP}, \delta_{CP}$) of the convergent point (see
the schematic representation in Fig.~\ref{Fig9}). Spherical
trigonometry tells us that
\begin{equation}\label{lambdadef}
\cos \lambda = \sin \delta \sin \delta_{CP} + \cos \delta \cos
\delta_{CP} \cos(\alpha_{CP}-\alpha).
\end{equation}

If we now change coordinates (see Fig.~\ref{Fig9}) and project the
proper motion onto the axis joining the star to the CP, on one hand
(component $\mu_\parallel$), and onto the perpendicular direction,
on the other (component $\mu_\perp$), the proper motion components
in both systems are related by
\begin{equation}
\left\{ \begin{array}{l} \mu_\parallel = \sin \theta \: \mu_\alpha
\cos
\delta + \cos \theta \: \mu_\delta \\
\mu_\perp = -\cos \theta \: \mu_\alpha \cos \delta + \sin \theta \:
\mu_\delta.
\end{array}
\right.
\end{equation}
with
\begin{equation}
\tan \theta =
\frac{\sin(\alpha_{CP}-\alpha)}{\cos\delta\tan\delta_{CP} -
\sin\delta\cos(\alpha_{CP}-\alpha)}.
\end{equation}

If the proper motion vector is directed exactly toward the CP, then
we also have
\begin{equation}\label{muparaandperp}
\left\{ \begin{array}{l} \mu_\parallel = \frac{\pi v \sin
\lambda}{A} =
\left|\;\vec{\mu}\;\right| \\
\mu_\perp = 0.
\end{array}
\right.
\end{equation}
In general, the measurement errors $\sigma_\perp$ and velocity
dispersion within the moving group $\sigma_{int}$ (expressed in
mas/yr) conspire to make $\mu_\perp$ different from zero, but the
expectation value of this quantity for any star belonging to a
moving group is zero. If one assumes, following
\cite{1999MNRAS.306..381D}, that the error-weighted value of
$\mu_\perp$, defined as
\begin{equation}\label{tperp}
t_\perp \equiv \frac{\mu_\perp}{\sigma_{tot}}
\end{equation}
where
\begin{equation}
\sigma_{tot} \equiv \sqrt{\sigma_\perp^2+\sigma_{int}^2},
\end{equation}
is distributed normally with zero mean and unit variance for all
stars in the moving group, then the probability distribution for a
given combination of $\mu_\perp$ and $\sigma_{tot}$ to occur is
\begin{equation}\label{indivproba}
p_{ind} = \frac{1}{\sqrt{2\pi}}\exp(-\frac{t_\perp^2}{2}).
\end{equation}
The CP method for finding a moving group within a cluster of $N$
stellar candidates, as implemented by \cite{1971MNRAS.152..231J},
goes through the following steps \citep[see
also][]{1999MNRAS.306..381D}.
\begin{enumerate}
\item First define a grid $(i, j)$ in the plane of the sky $(\alpha, \delta)$,
where $i$ and $j$ vary from 1 to $N_{grid}$. The grid defines
candidate CP positions.
\item At each grid point $(i,j)$,  use Eq.~\ref{tperp} to compute the
expression
\begin{equation}\label{X2}
X^2 = \sum_{k=1}^N (t_\perp^2)_k
\end{equation}
where the index $k$ runs over all stars. With $t_\perp^2$
distributed normally, $X^2$ is distributed as $\chi^2$ with $N-2$
degrees of freedom. Minimizing $X^2$ is equivalent to maximizing the
likelihood that the computed set of $t_\perp$ occurs, and the most
likely CP is therefore the grid point with the lowest $X^2$ value.
\item However, it could still occur that the lowest
value of $X^2$ is due to chance rather than to a good fit between
observations and the model. To evaluate this possibility, one
calculates the probability $p_{min}$ for $X^2$ to be higher than its
computed value by chance even if the derived CP is the correct one.
This probability is given by the incomplete gamma function
\begin{equation} \label{pmin}
p_{min} = \frac{1}{\Gamma[\frac{1}{2}(N-2)]}\int_{X^2}^\infty
x^{\frac{1}{2}(N-2)-1}\,\exp(-x)\,dx.
\end{equation}
\item If $p_{min}$ is too low, then one rejects the star with the
highest $\left|\,t_\perp\,\right|$, corrects the number of stars and
goes back to step 2.
\item One then continues until $p_{min}$ has reached an acceptable value (to be
discussed later). When this is done, all remaining stars are defined
as \emph{bona fide} group members and the most likely CP candidate
in the last iteration is defined as the convergent point of the
group.
\end{enumerate}

As noted by \cite{1999MNRAS.306..381D}, the basic equations of the
method (Eqs.~\ref{BasicEq} and \ref{muparaandperp}) remain valid if
the stars belonging to the association are in a state of uniform
expansion. In that case, however, using the proper motion
information to determine the CP as we do here implies that the
derived velocity $\vec{v}$ is the sum of the actual space motion and
the reflex motion of the expansion, and these two motions cannot be
disentangled.

Another caveat with the above method is its bias towards low
$t_\perp$, which is unavoidable since on principle the CP search
favors low values of this parameter. However, not all stars with low
$t_\perp$ are necessarily moving group members; both stars with
intrinsically small proper motions and stars located at large
distances have a low $t_\perp$.

It is therefore customary to eliminate stars with insignificant
proper motions prior to doing the analysis, because such objects
would not be rejected in the CP search. The criterion for rejection
\citep{1999MNRAS.306..381D} is
\begin{equation} \label{tmin}
t = \frac{\mu}{\sigma_\mu} = \frac {\sqrt{\mu_\alpha^2 \cos^2\delta
+\mu_\delta^2} }{ \sqrt{\sigma^2_{\mu_{\alpha \cos \theta}}
+\sigma^2_{\mu_\delta}}} \leq t_{min}
\end{equation}
where $t_{min}$ depends on the quality of the data (see below).
While this preliminary step eliminates a number of non-group
members, it does nothing to correct the bias toward selecting the
most distant stars in the group as members.

\subsection{Modified CP method}

The main difficulty that we encountered while using this method of
finding a moving group of stars in Taurus is due to a combination of
two properties of this region. First, the Taurus-Auriga star-forming
region spans a large volume, but the YSOs are bunched in small
groups with vast spaces devoid of stars between them rather than
being distributed more evenly on the plane of the sky, as, e.g.,
older open clusters are. This specific morphology complicates the
search for the CP, since the most likely CP position wanders in the
plane of the sky depending on the groups of stars that dominate
$X^2$ during a given iteration. As a consequence, a large number of
possible group members may well be eliminated before $p_{min}$
reaches a reasonable value. Moreover, the remaining stars are the
most distant ones, as explained above.

Second, there is a velocity dispersion between the different parts
of the Taurus star-forming region of several km/s, which was first
noted by \cite{1979AJ.....84.1872J} and confirmed by subsequent
investigators. Although this velocity dispersion is taken explicitly
into account in the CP method, it reduces both the computed
individual probabilities $p_{ind}$ (Eq.~\ref{indivproba}) of stars
being members of a moving group and the likelihood that the computed
set of $t_\perp$ occurs, thus making it more difficult to reach a
high $p_{min}$. Extensive Monte-Carlo simulations (cf.
Sect.~\ref{MCSims}), which assumed a group of stars with the same
positions as our sample and various sets of parameters, indeed
confirmed that both the size of the recovered moving group and the
probability of recovering the correct average distance for that
moving group decreases with increasing internal velocity dispersion.

One possibility for considering the specificities of our sample
would be to use the CP search procedure as described above while
lowering the requirement on $p_{min}$ to a value in the range
$10^{-1}$. However, we felt uncomfortable with settling for low
$p_{min}$ values that do not confirm the reality of the derived
moving group. For comparison, \cite{1999MNRAS.306..381D} chose
$p_{min}$ equal to 0.954 in his study of moving groups, also on the
basis of Monte-Carlo simulations.

Instead, we devised a modification of the CP method that helped us
to find the moving group while giving very low probabilities that
the solution was due to chance. The idea is to provide some initial
guidance to the code by first guessing the CP position (exploiting
the available radial velocity data) and eliminating those stars with
the most obviously discrepant proper motions before proceeding with
the classic CP search as described above.

For the initial guess of the CP position, we used the sub-sample of
stars with known radial velocities $v_{rad}$ and the average,
post-Hipparcos Taurus-Auriga parallax value derived by
\cite{1999A&A...352..574B}. Starting from Eq.~\ref{vtan}, which
defines the two tangential components $v_\alpha$ and $v_\delta$, we
converted the velocity components in the equatorial system to a
rectangular coordinate system $(x,y,z)$ in which $x$ points towards
the vernal equinox, $y$ is directed towards the point on the equator
with $\alpha = \pi/2$, and $z$ points towards the Northern
equatorial pole. The conversion is done by the following
transformation
\begin{equation}
\left( \begin{array}{c} v_x \\ v_y \\ v_z \end{array} \right) =
\left( \begin{array}{ccc} -\sin\alpha & -\sin\delta\,\cos\alpha &
\cos\delta\;\cos\alpha \\
\cos\alpha & -\sin\delta\;\sin\alpha & \cos\delta\;\sin\alpha \\
0 & \cos\delta & \sin\delta \end{array} \right) \left(
\begin{array}{c} v_\alpha \\ v_\delta \\ v_{rad} \end{array}
\right).
\end{equation}
An approximate CP location is then given by
\begin{equation}
\left\{ \begin{array}{lll} \alpha_{CP} & = & \arctan
\left(\frac{\overline{\textstyle v_y}}{\overline{\textstyle v_x}}\right) \\ & & \\
\delta_{CP} & = & \arctan \left(\frac{\overline{\textstyle
v_z}}{\left(\overline{\textstyle v_x}^2 + \overline{\textstyle
v_y}^2\right)^{1/2}}\right)
\end{array} \right.
\end{equation}
where $\overline{v_x}$, $\overline{v_y}$, and $\overline{v_z}$ are
the averages of the velocity components over the sample of stars
with known radial velocities.

Once this first guess of the CP coordinates was completed, we
computed the individual probability of each star being part of a
moving group converging to this point by using Eq.~\ref{indivproba}
and eliminated those objects for which $p_{ind}$ was lower than a
preset value.

We then proceeded with the usual CP method as described above.
Monte-Carlo simulations with various sets of parameters reported in
Sect.~\ref{MCSims} show that the convergence to a highly probable
solution is very quick after the initial elimination of discrepant
stars. Note also that the overall probability that the group is not
due to chance is generally very close to one, and much higher when
we perform such an initial screening than when using the usual CP
method, which confirms that this simple procedure eliminates those
interlopers whose proper motions are not compatible with the derived
CP. Conversely, and this is the main drawback of the modified
method, more \emph{bona fide} group members are usually eliminated
than when using the usual CP method, depending on the choice of
computational parameters.

We should not expect this modified CP search method to produce
results that are different from the classic method -- both methods
rely in essence on the same procedure -- unless the initial CP guess
biases the final result. This can be checked because the second step
of the search allows for an iteration of the CP coordinates. In the
Monte-Carlo simulations discussed in Sect.~\ref{MCSims}, the final
CP coordinates are usually close to the coordinates of the initial
guess, thus confirming the lack of appreciable bias. We emphasize
that the only advantage to this search strategy over the classic one
for the stellar group under investigation is that it often leads to
a converged solution, whereas the classic method fails to converge
due to the specific properties of Taurus-Auriga discussed above.

\subsection{Parallax computations}

Once we have defined a moving group, we can derive the kinematic
parallaxes $\pi_{Vrad}$ of individual group members and their
uncertainties, if we know their radial velocities $V_{rad}$. We have
\begin{equation}
\pi_{Vrad} = \frac{A \mu_\parallel}{V_{rad} \tan \lambda}
\end{equation}
where $\lambda$ is given by Eq.~\ref{lambdadef}. The error on
$\lambda$ is computed as in \cite{1999MNRAS.306..381D}, and standard
error propagation techniques lead to a determination of the
uncertainty on $\pi$.

In order to estimate approximate individual parallaxes of group
members with unknown radial velocities, we derive the average
spatial velocity $V_{group}$ from the Galactic velocity components
of the stars with known radial velocities and use the relationship
\begin{equation}
\pi_{Vgroup} = \frac{A \mu_\parallel}{V_{group} \sin \lambda}.
\end{equation}
Again, one can derive the uncertainty on $\pi$ from the uncertainty
on the group velocity, $\mu_\parallel$ and $\lambda$. A comparison
of both parallax estimates is made in the section describing the
results for Taurus-Auriga (Sect.~\ref{KinematicAnalysisSection}).

This ends the general description of the method used to find the
moving group members and their individual parallaxes. In the
following section, we present the extensive Monte-Carlo simulations
that were used to test our alternate CP method and compare it to the
classic one.

\begin{figure*}
\sidecaption
\includegraphics[width=12cm]{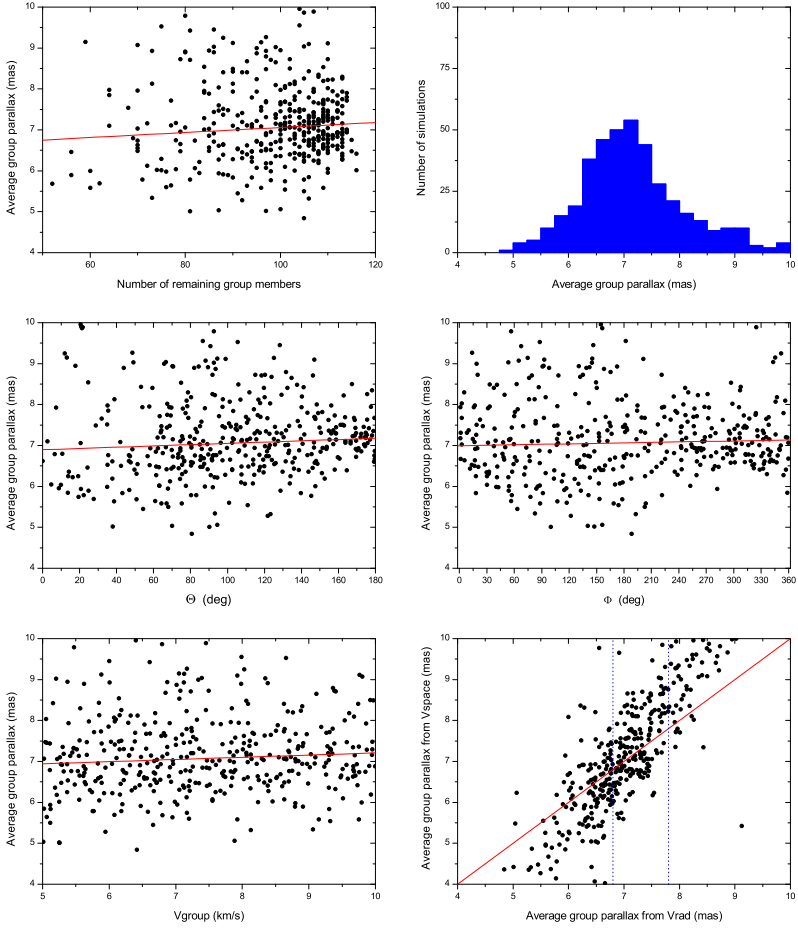}
\caption[]{Results of the classic CP search for Simulation 1, the
parameters of which are given in Table~\ref{MCResults}. Upper left
panel: relationship between the average group parallax and the
number of group members identified by the CP search for each
realization. Upper right panel: histogram of recovered parallaxes.
Middle left panel: average parallax as a function of the angle
$\Theta$. Middle right panel: average group parallax as a function
of the angle $\Phi$. Lower left panel: average parallax as a
function of the initial group velocity. Lower right panel:
comparison of the recovered parallaxes using either the observed
radial velocities of individual stars or the average space velocity
of the group, as discussed in the text. Linear, error-weighted fits
to the data points are also shown in each panel, except for the
lower right one, where for the purpose of comparison we have drawn a
straight line indicating perfectly correlated parallaxes. The two
vertical lines at $\pi = 6.3$ and $\pi = 7.8$ indicate the $\pm
1\sigma$ range of parallaxes assigned to synthetic stars in the
simulations.} \label{Fig10}
\end{figure*}

\begin{figure*}
\sidecaption
\includegraphics[width=12cm]{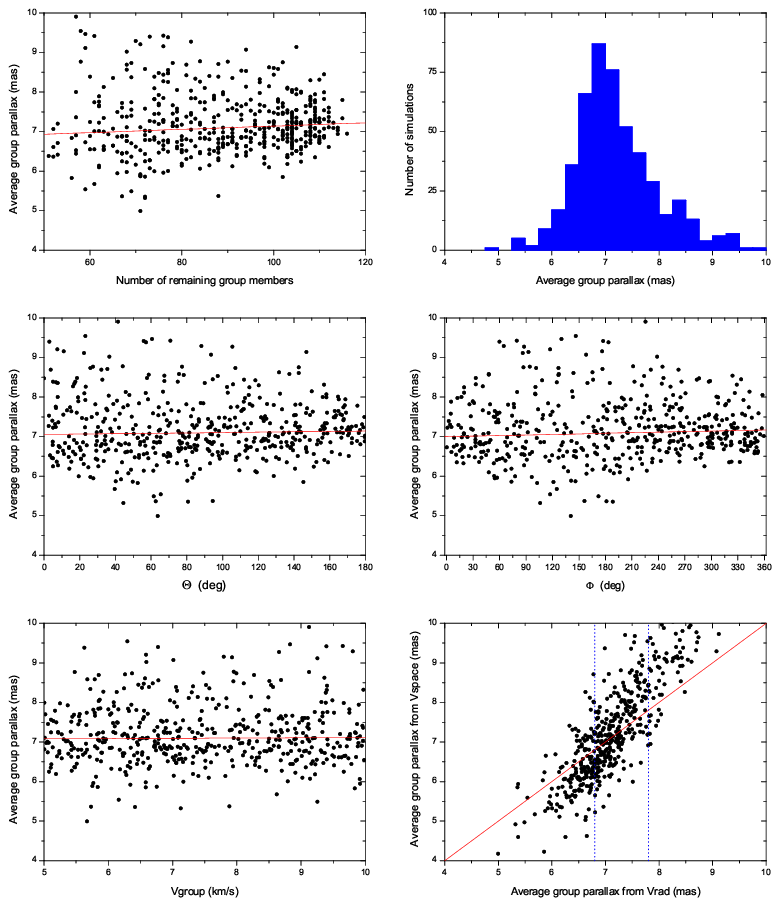}
\caption[]{Results of the variant CP search for Simulation 5. See
the caption of Fig.~\ref{Fig10} for details.} \label{Fig11}
\end{figure*}

\section{Monte-Carlo simulations of the moving group
search}\label{MCSims}

Monte-Carlo simulations are often useful for exploring the role of
the various parameters in numerical methods such as the CP search.
They are also crucial for gaining confidence in the derived results.

J. \cite{1999MNRAS.306..381D} performed extensive simulations that
convincingly demonstrate the ability of the CP method to find and
eliminate interlopers, i.e., field stars that are seen projected
against the stellar group but are not members of the moving group.
The method also eliminates stars from the moving group that are
actual members, and the computation parameters are chosen so as to
find the largest number of likely group members.

In our simulations we focused mainly on how the classic CP method
and its alternative deal with the large internal velocity dispersion
found in our stellar sample, as well as with the relatively large
uncertainties in the measured proper motions. These two aspects were
not discussed by \cite{1999MNRAS.306..381D} since he was dealing
with groups having low internal velocity dispersion and using
accurate Hipparcos proper motions. We also investigated how
interlopers affect our results.

\subsection{Monte-Carlo simulation set-up}

In the following, we compare results obtained using the same data
sets by the classic CP method, on one hand, and by its variant with
an initial guess of the CP coordinates, on the other.

We constructed our synthetic groups of stars in the same way as
\cite{1999MNRAS.306..381D}. The total number of stars that we used
in these simulations was 117, corresponding to the sample common to
both the \cite{2005A&A...438..769D} and \cite{1988cels.book.....H}
catalogs. This choice of  sample is justified in
Sect.~\ref{KinematicAnalysisSection}. Each synthetic star had the
same equatorial coordinates as one of our program stars.  For each
realization, we randomly chose a group spatial velocity that is
shared by all members of the data set. This was done by randomly
drawing the velocity modulus in a given interval (typically between
5 and 10 km/s), as well as the two angles $\Theta$ and $\Phi$
defining its direction in the $(X,Y,Z)$ coordinate system. We note
in passing that the approximate values of these parameters for the
Taurus-Auriga group discussed in
Sect.~\ref{KinematicAnalysisSection} are $V_{group} = 7.3$ km/s,
$\Theta = 116^\circ$, and $\Phi = 171^\circ$. We have neglected in
this derivation the small correction accounting for the difference
between the Galactic rotation values for the Sun and the Taurus
region.

In some simulations, we injected a certain percentage of field stars
(between 3 and 10\% -- see discussion in the following section) in
order to test the response of the code to their presence. This was
done randomly and the proper motion values were assigned by randomly
drawing a star from the sample of Hipparcos stars described in
Sect.~\ref{CatalogDescriptionSection}.  Since Hipparcos measurement
uncertainties are lower than in our sample, we added measurement
errors to these data using the procedure described below for moving
group members. If the coordinates of the simulated interloper
corresponded to a catalog star with known radial velocity, we
assigned the observed value of the radial velocity and its
uncertainty to the interloper.

Otherwise, stars were considered to be members of the moving group
and we thus computed the expected values of the Galactic velocity
components from the assumptions described above. We then added a
randomly drawn velocity dispersion to each velocity component. These
internal motion components were normally distributed with width
$\sigma_{int}$. In many simulations, we used $\sigma_{int} \approx
6$~km/s, which is representative of the velocity dispersion of our
stellar sample as found in proper motion surveys. The Taurus
star-forming region is made up of subgroups of stars, the members of
which are sometimes separated by only a few tenths of a parsec in
projection. In these subgroups, the velocity dispersion of
individual stars is much smaller than the dispersion between the
velocities of various subgroups \citep[cf.][]{1979AJ.....84.1872J}.
In the simulations, we therefore assumed that velocities of stars
closer than 1 pc in projection were the same except for measurement
errors.

Once the velocity dispersion was added to the Galactic velocity
components, the values $(U,V,W)$ were corrected for the solar
motion, using $(U,V,W)_\odot\footnote{There is a large array of
post-Hipparcos values for the solar motion [see the discussion in
\cite{1998MNRAS.298..387D}], and we chose the values derived from
K0-K5 giants here because our PMS sample includes many stars in this
spectral range, although they are usually luminosity class IV-V
rather than III. These values are not widely different from the
``classical" solar motion values given, e.g., by
\cite{1968gaas.book.....M}.} = (9.88,14.19,7.76)$~km/s
\citep{2000A&A...354..522M}. The stellar parallax was then drawn
from a normal distribution centered on the average parallax of
Taurus with width equal to the standard error computed by
\cite{1999A&A...352..574B}: $\pi_{Taurus} = 7.31 \pm 0.49$~mas.

From there, proper motions and radial velocities were computed and
measurement errors added to the results. In the simulations reported
here, we drew random errors from a normal distribution with width
equal to the average uncertainties of the observed proper motions
and radial velocities. In other simulations not illustrated here, we
used the observational uncertainties given in the
\cite{2005A&A...438..769D} catalog for the members of the moving
group. The simulation results were similar to those reported here.
Note that we assigned a radial velocity value only to those objects
that simulated stars with known radial velocity in our data set.
Finally, we added the Galactic rotation using the first-order
formulae and the same Oort constants as \cite{1999MNRAS.306..381D}.

After constructing such a synthetic data set, we ran the CP search
and recorded various simulation parameters $(v_{group}, \Theta,
\Phi)$ and some of the results, notably the number of group members
recovered by the search and the average group parallax, which can be
directly compared to the known input parameters.

\subsection{Results}

\begin{table}
\caption{Parameters and main results of Monte-Carlo
simulations}\label{MCResults}
\begin{center}
\tiny{
\begin{tabular}{lccccccc}
\hline\hline
Simulation &  1 &  2 & 3 & 4 & 5 & 6 & 7 \\
Figure & \ref{Fig10} & - & - & \ref{Fig11} & - & - & - \\
\hline
\multicolumn{8}{c}{Parameters} \\
\hline
$\sigma_{dis}$ & 6 & 6 & 6 & 6 & 6 & 6 & 3  \\
$\sigma_{PM}$ & 5 & 5 & 5 & 5 & 5 & 5 & 2 \\
$N_{intl}/N_{total}$ & 0 & 0 & 0 & 0 & 0.03 & 0.1 & 0 \\
$1-p_{min}$ & $10^{-1}$ & $10^{-5}$ & $10^{-5}$ & $10^{-10}$ & $10^{-10}$ & $10^{-10}$ & $10^{-10}$ \\
$p_{ind}$ & - & 0.6 & 0.7 & 0.8 & 0.8 & 0.8 & 0.7 \\
\hline
\multicolumn{8}{c}{Results} \\
\hline
$p_\pi$ & 0.32 & 0.44 & 0.44 & 0.51 & 0.49 & 0.48  & 0.64 \\
$f$ & 0.18 & $< 0.02$ & $< 0.02$ & $< 0.02$ & $< 0.02$ & $< 0.02$ & $< 0.01$ \\
$\bar{N}/N_{total}$ & 0.84 & 0.84 & 0.81 & 0.75 & 0.73 & 0.69 & 0.91 \\
\hline
\end{tabular}
}
\end{center}
\end{table}

We constructed 500 such data-set realizations when running one
Monte-Carlo simulation for a given set of CP-search parameters. Some
results are reported below.

Table~\ref{MCResults} gives the main parameters of the simulation,
i.e., the internal velocity dispersion $\sigma_{dis}$ in km/s, the
average uncertainty of proper motions measurements $\sigma_{PM}$ in
mas/yr, the requested minimum probability that the group is not due
to chance, as well as the minimum individual probability of a star
being a group member (see Sect.~\ref{CPMethodSection}). When the
classic CP method is used, this parameter is not needed, and the
entry is marked as $-$. The simulation results, also summarized in
the table, are the probability $p_{\pi}$ of recovering the parallax
within the assumed error bars for a given set of simulations, the
percentage $f$ of cases for which the moving group was not detected,
and the average number of recovered group members in percentage
$\bar{N}/N_{total}$.

The main results are illustrated by two figures. Figure~\ref{Fig10}
shows results of the classic CP search for Simulation 1, and
Fig.~\ref{Fig11} shows a set of results obtained using the alternate
CP search method, with parameters close to the ones we chose for the
actual computations. These two figures have six panels each, and
their contents are described in the caption of Fig.~\ref{Fig10}.

Simulation 1 implements the classic CP method for group parameters
similar to those of our data set. Table~\ref{MCResults} shows that
the probability of recovering the true parallax of the moving group
is only 0.32, and the failure rate $f$ of detecting the moving group
is as high as 18\%, whereas 84\% of the moving group members are
recovered when using this method.

Simulations 2 to 7 implement the modified CP method discussed in
Sect.~\ref{CPMethodSection}. Results are given for three different
values of the threshold value for the individual probabilities
$p_{ind}$ of a star being a group member, for two values of the
$(\sigma_{disp}, \sigma_{PM})$ parameter couple, and for three
values of $N_{intl}/N_{total}$, the fraction of interlopers among
the sample of stars. The results obtained with $p_{ind} = 0.6$
(Simulation 2) roughly correspond to those of the classic CP method
in the sense that 84\% of the group members are recovered, but the
failure rate for finding a moving group is much lower ($<$ 2\%), and
the probability of recovering the right average parallax is 44\%.

As expected, the number of recovered group members decreases with
increasing $p_{ind}$, while the probability of recovering the true
parallax reaches about 0.5 for $p_{ind} \approx 0.8$.

We should note here that the total number of stars in the sample has
little influence on the results, as long as it is sufficiently large
for this statistical search method to be meaningful. In addition to
the simulations reported here, which used 117 stars, we performed
simulations for the total sample of 217 stars. Results with this
larger group differed little from those reported in this section,
although the probabilities of recovering the right average parallax
value were higher by a few \%.

One also notes that the probability of detecting the moving group
with this method is always very high once  $p_{ind} > 0.6$. The
method appears tolerant of an interloper fraction up to 10\%
(Simulations 5 and 6), which barely affects the results. In all
cases we investigated, about half of the interlopers were rejected
as non-members of the moving group, while the second half remained
undetected.

The last simulation considers lower values of the internal velocity
dispersion and proper motion measurement errors, and confirmed that
the CP search is much more accurate when high quality data are
available and the internal velocity dispersion is small. We should
nevertheless emphasize here that there is no need to use this
alternate CP search method when the internal velocity dispersion and
proper motion measurement uncertainties are as low as in this last
simulation; the classic CP search method works very well in such
cases.

There is no apparent correlation between the average parallax
derived from the CP search and the group velocity for the range of
velocities (5 to 10 km/s) that we investigated here. For these low
group velocities, there is a correlation between the derived
parallax and the angles $\Theta$ and $\Phi$, in the sense that the
dispersion of recovered parallaxes is smallest for in range of
angles for which the contrast between reflex solar motion and group
motion is highest.

\begin{table}
\caption{Taurus-Auriga stars with insignificant proper
motions}\label{InsignificantPM}
\begin{center}
\tiny{
\begin{tabular}{lc}
\hline\hline Star        &   HBC   \\
\hline
NTTS 040012+2545 AB &   357 \\
NTTS 040142+2150SW  &   360 \\
NTTS 040142+2150NE  &   361 \\
DH   Tau    &   38  \\
V710 Tau AB &   51  \\
CoKu HP Tau G3  &   414 \\
CoKu Tau-Aur Star 4 &   421 \\
DQ   Tau    &   72  \\
Haro 6-37 AB    &   73  \\
StHA   34   &   425 \\
V836 Tau    &   429 \\
\hline
\end{tabular}
}
\end{center}
\end{table}

Results of the alternate CP search method are sensitive to the
threshold probability $p_{ind}$, which must be carefully chosen to
achieve the best possible compromise between $p_{min}$ and the
recovered number of group members. Extensive tests suggest an
optimal threshold value in the range $0.7 - 0.8$ for our data set.

Parallaxes are computed in two ways. First, we use the radial
velocity information to compute parallaxes for the subset of stars
for which this quantity is known. Once this is done, we compute the
average group velocity for this subset and then assume that it is
the group velocity of all stars in the moving group. Approximate
parallaxes can then be computed for all stars in the moving group.
We compared the average moving group parallaxes obtained by these
two methods (see Figs.~\ref{Fig10} and \ref{Fig11}), and conclude
that they give similar results within the computed uncertainties
when the recovered average parallaxes are in the correct range
(delimited by two vertical lines in the figures). The general bias
of the CP method toward lower parallaxes that we discussed earlier
is apparent in these figures.

\subsection{Conclusion}

We have presented an alternate CP search method suitable for
dispersed stellar groups with large internal velocity dispersions.
For such groups, the classic CP method often fails to converge or to
give realistic results. The alternate method is much less
computationally intensive and nearly always converges when a moving
group is present, provided the fraction of interlopers is not too
large. However, it tends to eliminate more actual group members than
the classic method does. Its use should therefore be limited to
cases where the classic method is expected to fail.

\begin{figure*}
\resizebox{\hsize}{!} {\includegraphics[angle=0]{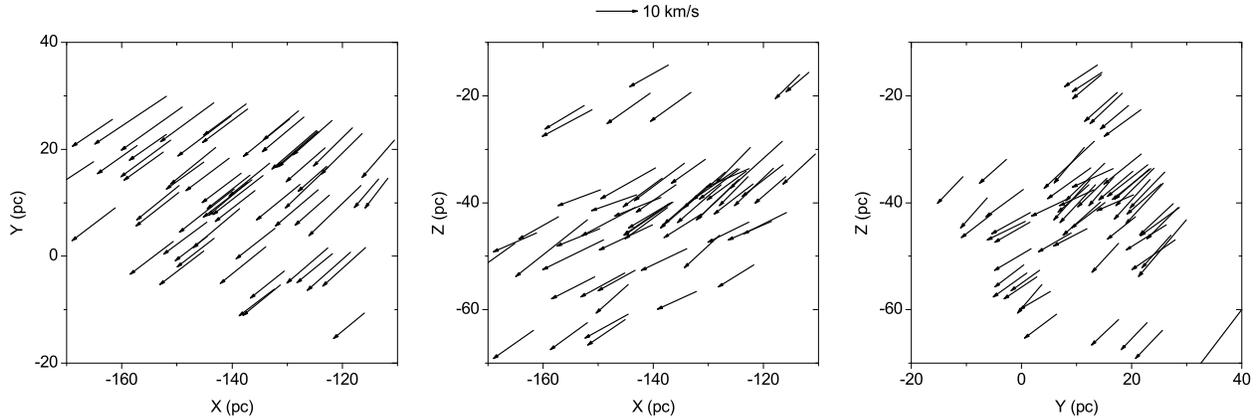}}
\caption[]{Projection on the 3 planes XY, XZ, and YZ of the spatial
velocities for the Taurus moving group members with known radial
velocities. The velocity vectors originate at the stars' coordinates
in the XYZ reference frame.} \label{Fig12}
\end{figure*}

\begin{figure*}
\sidecaption
\includegraphics[width=12cm]{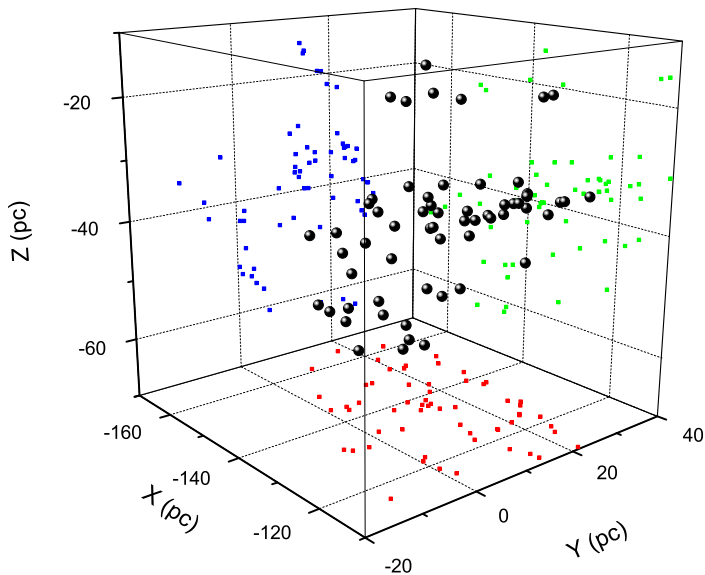}
\caption[]{3D spatial distribution of the Taurus moving group
members with known radial velocities. The ball-like symbols
represent individual stars whose projections on the 3 planes are
shown as small squares with different colors.
Table~\ref{GroupGalVel} gives the Galactic coordinates and
velocities of individual objects.} \label{Fig13}
\end{figure*}

\section{Application to Taurus-Auriga}\label{KinematicAnalysisSection}

As discussed in Sect.~\ref{CatalogDescriptionSection}, we should
expect that an undefined, but probably significant, number of field
stars are not distinguishable, as far as their kinematics are
concerned, from the PMS stars that we are interested in. A
consequence, demonstrated by the simulations reported in
Sect.~\ref{MCSims}, is that up to about 50\% of those interlopers
may remain undetected even though a moving group has been identified
and hence may pollute the results. The best way to proceed is thus
to preemptively screen out the main sequence stars that may possibly
be present in our sample before searching for a moving group among
the young stellar population. A critical discussion of the PMS
status of the stars included in the \cite{2005A&A...438..769D}
catalog is thus in order. This is done in the next subsection, where
we divide the Taurus-Auriga sample into two subsets, one for stars
that are clearly of PMS nature, and the other for stars in a more
uncertain evolutionary state. We then discuss the values chosen for
the parameters that enter the computation, present the results of
the kinematic analysis for both subsets of stars, and give the list
of moving group members with their individual parallaxes.

\subsection{Refining the PMS star sample}

In their proper motion catalog for PMS stars,
\cite{2005A&A...438..769D} include a flag indicating the
\emph{``classification as given by CDS or found in the literature :
T=T Tauri (CTT or WTT), A = HAeBe, Y = YSO, W = WTT, P = PMS, X =
X-ray active source (Li 2000, table 2), O = Post T Tauri, L =
Emission Line from HBC catalogue, Tc = T Tauri candidate, Wc = WTTs
candidate. Ac = HAeBe candidate, Yc = YSOs candidate, Pc = PMS
candidate".} It is already apparent from this list that the catalog
might well be contaminated by main sequence stars, since many X-ray
active main sequence stars exist and also since the evolutionary
status of the various ``candidates" above is unclear. A closer look
at the list of catalogued objects unveils an even more serious
problem because most X-ray sources that are detected in the general
direction of Taurus are noted as PMS objects, although there is a
large body of evidence suggesting that they are, at least in part,
active main-sequence stars \citep[e.g., ][]{1992AJ....104..762G,
1997AJ....113..740B, 1997A&A...325..647N, 2000A&A...359..181W}.
Depending on what fraction of the stars detected through their X-ray
emission are truly PMS, there could be a sizable number (perhaps up
to 30\%) of field stars in the sample of ``PMS" stars discussed in
Sect.~\ref{CatalogDescriptionSection}.

We devised the following strategy to circumvent the problem created
by the presence of potential field stars in our sample.
\begin{enumerate}
    \item We chose a sub-sample of highly probable PMS objects by
restricting the \cite{2005A&A...438..769D} sample to stars contained
in the \cite{1988cels.book.....H} catalog of confirmed Orion
population members and ran the CP search for that group of 117
objects, thus looking for a \emph{core} Taurus-Auriga moving group.
    \item Using the derived CP coordinates, we computed the probability
of each star in the full sample of 217 stars being a member of the
core moving group and defined as possible additional members those
stars whose probability of membership was sufficiently high.
    \item We then verified, from a thorough literature search, the PMS
status of stars belonging to this extended moving group in order to
eliminate the possibly remaining interlopers.
\end{enumerate}
The details of the process are given below.

\subsection{Choice of computation parameters}

There are five such parameters: the number $N_{grid}$ of trial grid
points for the CP search, the value $t_{min}$ that defines the
magnitude of insignificant proper motions (see Eq.~\ref{tmin}), the
velocity dispersion $\sigma_{int}$ that determines the basic
quantity $t_\perp$ (Eq.~\ref{tperp}), the threshold probability
$p_{min}$ (Eq.~\ref{pmin}) for deciding that the moving group is not
a chance occurrence, and the threshold probability $p_{ind}$
(Eq.~\ref{indivproba}) of individual stars being considered as
possible moving group members in the initial screening described
above.

\begin{figure}
\resizebox{\hsize}{!}{\includegraphics[angle=0]{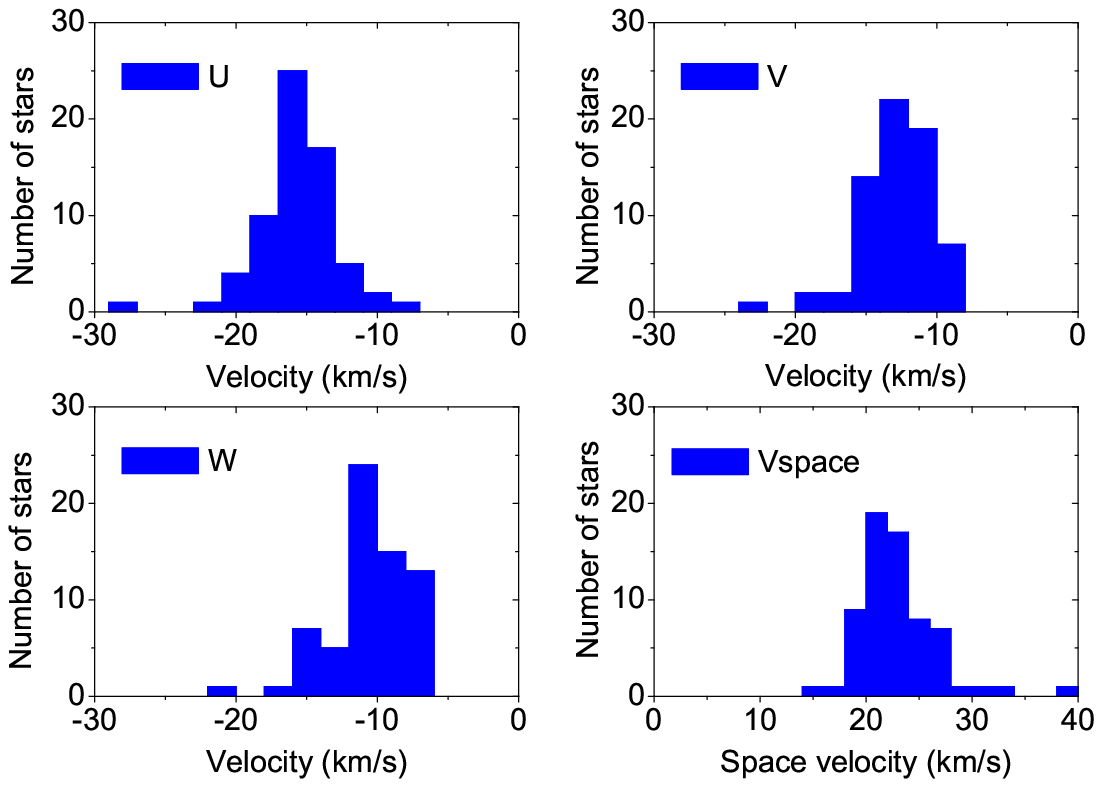}}
\caption[]{Histograms of the U, V, W galactic velocity components
and spatial velocity for members of the Taurus moving group with
known radial velocities.} \label{Fig14}
\end{figure}

\renewcommand{\arraystretch}{1.2}
\begin{table}
\caption{Parallaxes (in mas) and distances (in pc) of 58 moving
group stars with known radial velocities.}\label{VradPar}
\begin{center}
\tiny{
\begin{tabular}{lcrr}
\hline\hline Star &   HBC  &  $\pi_{Vrad} \pm
\sigma_\pi$ & $d_{Vrad} \pm \sigma_d$ \\
\hline
NTTS 035120+3154SW  &   352 &   3.86    $\pm$   0.74    &   259 $^{+    61  }_{ -42 }$\\
NTTS 035120+3154NE  &   353 &   4.44    $\pm$   0.78    &   225 $^{+    48  }_{ -34 }$\\
NTTS 040047+2603E   &   359 &   7.21    $\pm$   2.18    &   139 $^{+    60  }_{ -32 }$\\
V773 Tau    &   367 &   6.84    $\pm$   1.07    &   146 $^{+    27  }_{ -20 }$\\
CW   Tau    &   25  &   7.71    $\pm$   1.95    &   130 $^{+    44  }_{ -26 }$\\
FP   Tau    &   26  &   6.30    $\pm$   1.20    &   159 $^{+    37  }_{ -25 }$\\
CX   Tau    &   27  &   6.70    $\pm$   1.16    &   149 $^{+    31  }_{ -22 }$\\
LkCa 4  &   370 &   6.87    $\pm$   1.86    &   146 $^{+    54  }_{ -31 }$\\
CY   Tau    &   28  &   6.60    $\pm$   0.63    &   152 $^{+    16  }_{ -13 }$\\
LkCa 5  &   371 &   7.99    $\pm$   1.62    &   125 $^{+    32  }_{ -21 }$\\
NTTS 041529+1652    &   372 &   6.95    $\pm$   2.64    &   144 $^{+    88  }_{ -40 }$\\
V410 Tau    &   29  &   7.30    $\pm$   0.88    &   137 $^{+    19  }_{ -15 }$\\
DD   Tau    &   30  &   6.22:    $\pm$   1.22    &   161: $^{+    39  }_{ -26 }$\\
CZ   Tau    &   31  &   4.56:    $\pm$   0.81    &   219: $^{+    47  }_{ -33 }$\\
Hubble 4   &   374 &   8.12    $\pm$   1.50    &   123 $^{+    28  }_{ -19 }$\\
NTTS 041559+1716    &   376 &   6.37    $\pm$   2.50    &   157 $^{+    101 }_{ -44 }$\\
BP   Tau    &   32  &   7.46    $\pm$   0.66    &   134 $^{+    13  }_{ -11 }$\\
V819 Tau    &   378 &   8.56    $\pm$   1.79    &   117 $^{+    31  }_{ -20 }$\\
LkCa 7  &   379 &   7.60    $\pm$   1.49    &   132 $^{+    32  }_{ -22 }$\\
RY   Tau    &   34  &   7.62    $\pm$   0.68    &   131 $^{+    13  }_{ -11 }$\\
HD   283572 &   380 &   7.64    $\pm$   1.05    &   131 $^{+    21  }_{ -16 }$\\
IP   Tau    &   385 &   7.74    $\pm$   1.71    &   129 $^{+    37  }_{ -23 }$\\
DF   Tau    &   36  &   7.76    $\pm$   1.29    &   129 $^{+    26  }_{ -18 }$\\
NTTS 042417+1744    &   388 &   6.46    $\pm$   1.02    &   155 $^{+    29  }_{ -21 }$\\
DI   Tau    &   39  &   6.69    $\pm$   2.03    &   149 $^{+    65  }_{ -35 }$\\
IQ   Tau    &   41  &   7.80    $\pm$   1.96    &   128 $^{+    43  }_{ -26 }$\\
UX   Tau AB &   43  &   6.55    $\pm$   0.99    &   153 $^{+    27  }_{ -20 }$\\
FX   Tau    &   44  &   6.98    $\pm$   1.76    &   143 $^{+    48  }_{ -29 }$\\
DK   Tau    &   45  &   6.09    $\pm$   0.86    &   164 $^{+    27  }_{ -20 }$\\
V927 Tau    &   47  &   7.00    $\pm$   1.82    &   143 $^{+    50  }_{ -30 }$\\
NTTS 042835+1700    &   392 &   9.41    $\pm$   2.69    &   106 $^{+    42  }_{ -24 }$\\
HK   Tau    &   48  &   5.79    $\pm$   2.31    &   173 $^{+    114 }_{ -49 }$\\
L    1551-51    &   397 &   7.02    $\pm$   0.82    &   143 $^{+    19  }_{ -15 }$\\
V827 Tau    &   399 &   6.25    $\pm$   1.18    &   160 $^{+    37  }_{ -25 }$\\
V826 Tau    &   400 &   8.12    $\pm$   0.84    &   123 $^{+    14  }_{ -11 }$\\
V928 Tau    &   398 &   7.49    $\pm$   1.87    &   133 $^{+    44  }_{ -27 }$\\
GG   Tau    &   54  &   7.92    $\pm$   0.85    &   126 $^{+    15  }_{ -12 }$\\
GH   Tau    &   55  &   7.69    $\pm$   1.83    &   130 $^{+    41  }_{ -25 }$\\
DM   Tau    &   62  &   7.68    $\pm$   2.80    &   130 $^{+    75  }_{ -35 }$\\
CI   Tau    &   61  &   6.38    $\pm$   2.28    &   157 $^{+    87  }_{ -41 }$\\
NTTS 043124+1824    &   407 &   2.78    $\pm$   0.86    &   360 $^{+    160 }_{ -85 }$\\
DN   Tau    &   65  &   7.08    $\pm$   0.82    &   141 $^{+    19  }_{ -15 }$\\
LkCa 14 &   417 &   6.96    $\pm$   0.82    &   144 $^{+    19  }_{ -15 }$\\
DO   Tau    &   67  &   7.19    $\pm$   1.96    &   139 $^{+    52  }_{ -30 }$\\
HV   Tau    &   418 &   8.28    $\pm$   1.20    &   121 $^{+    20  }_{ -15 }$\\
VY   Tau    &   68  &   6.93    $\pm$   2.06    &   144 $^{+    61  }_{ -33 }$\\
LkCa 15 &   419 &   5.96    $\pm$   0.84    &   168 $^{+    28  }_{ -21 }$\\
IW   Tau    &   420 &   6.46    $\pm$   1.53    &   155 $^{+    48  }_{ -30 }$\\
CoKu LkH$\alpha$332 G2    &   422 &   7.92    $\pm$   1.68    &   126 $^{+    34  }_{ -22 }$\\
GO   Tau    &   71  &   6.94    $\pm$   1.77    &   144 $^{+    49  }_{ -29 }$\\
DR   Tau    &   74  &   7.25    $\pm$   2.70    &   138 $^{+    82  }_{ -37 }$\\
UY   Aur    &   76  &   6.48    $\pm$   1.01    &   154 $^{+    28  }_{ -21 }$\\
GM   Aur    &   77  &   7.37    $\pm$   1.98    &   136 $^{+    50  }_{ -29 }$\\
LkCa 19 &   426 &   6.44    $\pm$   0.83    &   155 $^{+    23  }_{ -18 }$\\
AB   Aur    &   78  &   8.79    $\pm$   1.48    &   114 $^{+    23  }_{ -16 }$\\
SU   Aur    &   79  &   6.99    $\pm$   0.73    &   143 $^{+    17  }_{ -13 }$\\
NTTS 045251+3016    &   427 &   8.66    $\pm$   0.96    &   116 $^{+    14  }_{ -12 }$\\
RW   Aur AB &   80  &   7.21    $\pm$   0.82    &   139 $^{+    18  }_{ -14 }$\\
\hline
\end{tabular}
}
\end{center}
\end{table}

The number of grid points $N_{grid}$ defined over the plane of the
sky determines the accuracy of the CP derived position. We typically
used a $1000 \times 1000 \: (\alpha, \delta)$ grid for the final
models and a $500 \times 500 \: (\alpha, \delta)$ grid for the
Monte-Carlo simulations.

Rather than using $t_{min}$ as a truly free parameter, we estimated
its value from the data quality, following a remark made by
\cite{1999MNRAS.306..381D}, who noted that one should basically
reject all stars with proper motion $\mu$ smaller than $3
\sigma_\mu$, i.e.,

\begin{equation}
t_{min} \approx \frac{3
\sigma_\mu}{\sqrt{\sigma^2_{int}+\sigma^2_\mu}}.
\end{equation}

From the average proper motion error $\sigma_\mu = 6$ mas/yr and a
typical velocity dispersion $\sigma_{int} = 6$ km/s, we derived
$t_{min}= 1.66$ at the mean distance of Taurus. Using this cut-off
value, 11 among the 117 stars of our sample have insignificant
proper motions. Table~\ref{InsignificantPM} lists these stars, which
were thus eliminated from the moving group search. (see
Sect.~\ref{CPMethodSection}). Their number in the
\cite{1988cels.book.....H}  catalog (HBC) is also given.

In our test computations, we assumed the  velocity dispersion
$\sigma_{int}$ to be in the range 5 to 6 km/s, in agreement with the
value derived by \cite{1979AJ.....84.1872J} for the velocity
dispersion between subgroups of stars in the Taurus cloud. As
discussed by these authors, the internal velocity within the stellar
groupings themselves is much lower than this value. Note that the
proper motion uncertainty $\sigma_{\mu}$, which also enters
$t_\perp$, is given in the \cite{2005A&A...438..769D} catalog for
each star.

We tested a wide range of values for probability $p_{min}$, which
assesses the reality of the moving group. Even for values of
$p_{min}$ as high as 0.95, some objects with obviously discrepant
galactic velocities were found to pollute our results, so we
concluded that a value very close to 1 was needed given the
\cite{2005A&A...438..769D} measurement errors and the large internal
velocity dispersion in Taurus.

The threshold probability $p_{ind}$ required for individual stars to
be considered members of the moving group was typically set between
0.7 - 0.8. As shown by the Monte-Carlo simulations (see
Sect.~\ref{MCSims}), the value of this important parameter largely
determines both the final probability $p_{min}$ that the group is
not due to chance and the final size of the moving group.

\subsection{A core Taurus-Auriga moving group} \label{CoreGroup}

It is important to realize that the final size of a moving group is
not a fixed number but is instead determined by the final
realization probability that one decides to adopt. Our approach here
was conservative, as we chose to enforce $p_{min} \ge 1 - 10^{-10}$
in order to estimate the size of the Taurus moving group.

We adopted this conservative attitude in spite of the fact that it
minimizes the size of the moving group  because we are interested
less in the number of individual parallaxes that we derive than in
the credibility of the moving group. Because moving group members
share a common destiny, we will argue in an upcoming paper (Bertout
\& Siess, in preparation) that the moving group found here is more
homogeneous and more significant in a statistical sense than the
overall group of Taurus-Auriga YSOs that has been traditionally used
for investigations of the global properties of this region.

In this way, we defined a highly probable moving group of 83 stars
or stellar systems in Taurus-Auriga for which we could determine
individual parallaxes. The corresponding threshold value of
$p_{ind}$ is 0.78 for our sample. This procedure is likely to
eliminate some \emph{bona fide} moving group members, as the results
of Monte-Carlo simulations of Sect.~\ref{MCSims} show, but it does
define a core Taurus moving group that can be regarded as real with
a high degree of confidence.

The derived CP coordinates for the moving group are

\medskip
\noindent $\left\{ \begin{array}{lll}
\alpha_{CP} & = &  79.88^\circ \pm 0.01^\circ  \\
\delta_{CP} & = & -16.74^\circ \pm 0.02^\circ,
\end{array}
\right.$
\medskip

\noindent which confirms our suspicion while inspecting
Fig.~\ref{Fig1} that the proper motion vectors were often pointing
toward the lower left corner of the figure. Note that the high
accuracy of the derived CP coordinates was made possible by zooming
in on the region surrounding the CP (while keeping the same number
of grid points) once it was approximately located.

\renewcommand{\arraystretch}{1.0}
\begin{table*}
\caption{Positions and Galactic velocities of 67 confirmed members
the Taurus moving group with known radial velocities.}
\label{GroupGalVel}
\begin{center}
\tiny{
\begin{tabular}{lrrrrrrr}
\hline\hline Star &   HBC  &   $X$ (pc) &   $Y$ (pc) &   $Z$ (pc) & $U \pm \sigma_U$ (km/s) &   $V \pm  \sigma_V$ (km/s) &   $W \pm \sigma_W$ (km/s) \\
\hline
NTTS 035120+3154SW  &   352 &   -237    &   76  &   -73 &   -18.63  $\pm$   2.24    &   -9.47   $\pm$   3.74    &   -6.17   $\pm$   2.45    \\
NTTS 035120+3154NE  &   353 &   -206    &   66  &   -64 &   -16.86  $\pm$   1.71    &   -8.71   $\pm$   3.12    &   -7.92   $\pm$   2.19    \\
NTTS 040047+2603E   &   359 &   -128    &   27  &   -46 &   -13.29  $\pm$   2.39    &   -10.91  $\pm$   5.68    &   -12.26  $\pm$   4.41    \\
V773 Tau    &   367 &   -137    &   28  &   -41 &   -15.97  $\pm$   3.79    &   -11.83  $\pm$   2.82    &   -11.77  $\pm$   2.08    \\
CW   Tau    &   25  &   -122    &   25  &   -36 &   -14.75  $\pm$   1.91    &   -13.04  $\pm$   5.41    &   -11.36  $\pm$   4.01    \\
FP   Tau    &   26  &   -149    &   28  &   -47 &   -22.98  $\pm$   3.92    &   -16.39  $\pm$   4.91    &   -11.16  $\pm$   3.23    \\
CX   Tau    &   27  &   -140    &   26  &   -44 &   -20.08  $\pm$   1.76    &   -15.17  $\pm$   4.25    &   -9.52   $\pm$   2.82    \\
LkCa 4  &   370 &   -137    &   28  &   -40 &   -17.01  $\pm$   2.00    &   -12.68  $\pm$   5.92    &   -11.84  $\pm$   4.45    \\
CY   Tau    &   28  &   -143    &   29  &   -41 &   -20.16  $\pm$   1.51    &   -14.79  $\pm$   2.27    &   -10.59  $\pm$   1.53    \\
LkCa 5  &   371 &   -118    &   24  &   -34 &   -14.80  $\pm$   3.90    &   -14.38  $\pm$   4.62    &   -14.14  $\pm$   3.68    \\
NTTS 041529+1652    &   372 &   -132    &   5   &   -57 &   -14.80  $\pm$   2.13    &   -11.56  $\pm$   6.16    &   -6.67   $\pm$   3.81    \\
V410 Tau    &   29  &   -130    &   26  &   -36 &   -17.53  $\pm$   3.81    &   -14.16  $\pm$   2.58    &   -15.38  $\pm$   2.08    \\
DD   Tau    &   30  &   -152    &   30  &   -43 &   -27.10  $\pm$   5.81    &   -18.14  $\pm$   6.02    &   -21.47  $\pm$   4.85    \\
CZ   Tau    &   31  &   -207    &   41  &   -59 &   -42.45  $\pm$   5.92    &   -23.65  $\pm$   7.65    &   -30.84  $\pm$   6.23    \\
Hubble 4   &   374 &   -116    &   23  &   -33 &   -15.51  $\pm$   1.73    &   -15.27  $\pm$   4.41    &   -11.89  $\pm$   3.19    \\
NTTS 041559+1716    &   376 &   -145    &   6   &   -61 &   -16.30  $\pm$   4.08    &   -12.13  $\pm$   6.74    &   -8.98   $\pm$   4.47    \\
BP   Tau    &   32  &   -127    &   26  &   -34 &   -15.70  $\pm$   1.03    &   -13.12  $\pm$   1.91    &   -13.61  $\pm$   1.51    \\
V819 Tau    &   378 &   -111    &   22  &   -31 &   -12.29  $\pm$   3.89    &   -14.21  $\pm$   4.45    &   -11.41  $\pm$   3.35    \\
LkCa 7  &   379 &   -124    &   23  &   -36 &   -15.59  $\pm$   1.74    &   -13.59  $\pm$   4.47    &   -14.99  $\pm$   3.67    \\
RY   Tau    &   34  &   -125    &   24  &   -34 &   -16.87  $\pm$   1.48    &   -14.58  $\pm$   2.00    &   -11.65  $\pm$   1.43    \\
HD   283572 &   380 &   -124    &   23  &   -34 &   -15.16  $\pm$   3.82    &   -13.10  $\pm$   2.59    &   -11.35  $\pm$   1.89    \\
IP   Tau    &   385 &   -123    &   20  &   -34 &   -14.69  $\pm$   1.72    &   -12.89  $\pm$   4.53    &   -10.74  $\pm$   3.35    \\
DF   Tau    &   36  &   -123    &   17  &   -35 &   -10.75  $\pm$   3.83    &   -9.50   $\pm$   2.26    &   -10.97  $\pm$   2.05    \\
NTTS 042417+1744    &   388 &   -145    &   4   &   -55 &   -12.15  $\pm$   1.52    &   -9.12   $\pm$   2.10    &   -10.84  $\pm$   1.71    \\
DI   Tau    &   39  &   -143    &   20  &   -39 &   -17.08  $\pm$   1.97    &   -13.93  $\pm$   6.45    &   -6.04   $\pm$   4.17    \\
IQ   Tau    &   41  &   -123    &   17  &   -34 &   -16.16  $\pm$   1.83    &   -15.40  $\pm$   5.67    &   -7.07   $\pm$   3.62    \\
UX   Tau AB &   43  &   -143    &   3   &   -53 &   -14.04  $\pm$   1.50    &   -10.73  $\pm$   2.21    &   -7.71   $\pm$   1.50    \\
FX   Tau    &   44  &   -137    &   15  &   -40 &   -16.18  $\pm$   1.75    &   -12.64  $\pm$   4.96    &   -9.69   $\pm$   3.52    \\
DK   Tau    &   45  &   -157    &   21  &   -43 &   -15.08  $\pm$   1.51    &   -10.59  $\pm$   2.35    &   -8.47   $\pm$   1.68    \\
V927 Tau    &   47  &   -136    &   14  &   -40 &   -18.32  $\pm$   1.88    &   -14.10  $\pm$   5.82    &   -12.60  $\pm$   4.34    \\
NTTS 042835+1700    &   392 &   -99 &   0   &   -38 &   -13.76  $\pm$   1.94    &   -14.10  $\pm$   5.37    &   -10.41  $\pm$   3.59    \\
HK   Tau    &   48  &   -165    &   18  &   -48 &   -14.98  $\pm$   2.05    &   -10.01  $\pm$   6.75    &   -10.90  $\pm$   5.39    \\
L    1551-51    &   397 &   -134    &   2   &   -49 &   -17.23  $\pm$   1.48    &   -13.90  $\pm$   2.13    &   -8.15   $\pm$   1.38    \\
V827 Tau    &   399 &   -151    &   3   &   -54 &   -16.36  $\pm$   3.80    &   -12.36  $\pm$   2.83    &   -8.35   $\pm$   2.01    \\
V826 Tau    &   400 &   -116    &   2   &   -42 &   -16.21  $\pm$   1.47    &   -14.54  $\pm$   1.92    &   -8.08   $\pm$   1.23    \\
V928 Tau    &   398 &   -128    &   13  &   -37 &   -16.46  $\pm$   1.83    &   -13.47  $\pm$   5.38    &   -14.25  $\pm$   4.34    \\
GG   Tau    &   54  &   -119    &   1   &   -43 &   -16.29  $\pm$   1.47    &   -14.48  $\pm$   1.97    &   -6.95   $\pm$   1.24    \\
GH   Tau    &   55  &   -124    &   12  &   -36 &   -16.24  $\pm$   1.82    &   -13.74  $\pm$   5.21    &   -15.22  $\pm$   4.32    \\
DM   Tau    &   62  &   -123    &   2   &   -44 &   -15.41  $\pm$   2.03    &   -13.20  $\pm$   6.52    &   -7.59   $\pm$   4.16    \\
CI   Tau    &   61  &   -150    &   12  &   -45 &   -16.21  $\pm$   1.96    &   -12.64  $\pm$   6.66    &   -6.56   $\pm$   4.34    \\
NTTS 043124+1824    &   407 &   -340    &   4   &   -117    &   -15.10  $\pm$   1.72    &   -9.14   $\pm$   3.94    &   -13.02  $\pm$   3.01    \\
DN   Tau    &   65  &   -136    &   12  &   -37 &   -14.99  $\pm$   1.49    &   -11.65  $\pm$   2.02    &   -10.38  $\pm$   1.53    \\
LkCa 14 &   417 &   -138    &   15  &   -35 &   -14.69  $\pm$   1.49    &   -11.12  $\pm$   2.03    &   -10.77  $\pm$   1.60    \\
DO   Tau    &   67  &   -134    &   15  &   -33 &   -18.93  $\pm$   5.89    &   -14.77  $\pm$   6.10    &   -14.15  $\pm$   4.84    \\
HV   Tau    &   418 &   -117    &   13  &   -29 &   -19.75  $\pm$   1.63    &   -18.97  $\pm$   4.20    &   -17.82  $\pm$   3.36    \\
VY   Tau    &   68  &   -139    &   9   &   -39 &   -17.50  $\pm$   1.84    &   -14.70  $\pm$   6.23    &   -7.02   $\pm$   4.03    \\
LkCa 15 &   419 &   -161    &   9   &   -46 &   -16.38  $\pm$   1.51    &   -12.38  $\pm$   2.46    &   -7.09   $\pm$   1.62    \\
IW   Tau    &   420 &   -150    &   13  &   -38 &   -15.98  $\pm$   1.64    &   -12.95  $\pm$   4.46    &   -6.01   $\pm$   2.92    \\
CoKu LkH$\alpha$ 332 G2    &   422 &   -122    &   11  &   -30 &   -12.99  $\pm$   1.49    &   -11.36  $\pm$   3.58    &   -13.42  $\pm$   2.83    \\
GO   Tau    &   71  &   -140    &   13  &   -33 &   -20.67  $\pm$   4.01    &   -16.39  $\pm$   6.20    &   -11.65  $\pm$   4.44    \\
DR   Tau    &   74  &   -131    &   -5  &   -42 &   -14.22  $\pm$   1.97    &   -11.45  $\pm$   5.61    &   -8.94   $\pm$   4.03    \\
UY   Aur    &   76  &   -151    &   22  &   -23 &   -18.87  $\pm$   3.93    &   -13.91  $\pm$   3.00    &   -10.10  $\pm$   1.95    \\
GM   Aur    &   77  &   -133    &   17  &   -19 &   -15.33  $\pm$   1.66    &   -12.46  $\pm$   5.44    &   -10.87  $\pm$   4.49    \\
LkCa 19 &   426 &   -153    &   19  &   -22 &   -14.85  $\pm$   1.50    &   -10.58  $\pm$   2.17    &   -8.85   $\pm$   1.72    \\
AB   Aur    &   78  &   -112    &   15  &   -16 &   -8.61   $\pm$   4.92    &   -11.26  $\pm$   2.42    &   -7.22   $\pm$   1.64    \\
SU   Aur    &   79  &   -141    &   18  &   -19 &   -16.28  $\pm$   1.50    &   -12.19  $\pm$   2.00    &   -11.47  $\pm$   1.66    \\
NTTS 045251+3016    &   427 &   -113    &   14  &   -16 &   -9.26   $\pm$   1.49    &   -10.81  $\pm$   1.72    &   -9.08   $\pm$   1.40    \\
RW   Aur AB &   80  &   -137    &   14  &   -14 &   -14.53  $\pm$   1.50    &   -12.35  $\pm$   2.04    &   -8.24   $\pm$   1.53    \\
GSC  01262-00421    &   -   &   -152    &   23  &   -62 &   -13.97  $\pm$   1.51    &   -9.80   $\pm$   2.34    &   -10.21  $\pm$   1.69    \\
RX   J0405.7+2248   &   -   &   -162    &   26  &   -64 &   -15.21  $\pm$   1.95    &   -10.21  $\pm$   2.55    &   -10.71  $\pm$   1.86    \\
RX   J0406.7+2018   &   -   &   -145    &   18  &   -62 &   -14.09  $\pm$   1.93    &   -10.21  $\pm$   2.35    &   -9.63   $\pm$   1.68    \\
RX   J0423.7+1537   &   -   &   -122    &   0   &   -52 &   -13.34  $\pm$   1.91    &   -10.90  $\pm$   2.00    &   -8.33   $\pm$   1.43    \\
RX   J0432.8+1735   &   -   &   -145    &   1   &   -53 &   -16.61  $\pm$   2.43    &   -12.70  $\pm$   6.43    &   -8.98   $\pm$   4.29    \\
V1078 Tau  &   -   &   -244    &   2   &   -84 &   -18.01  $\pm$   1.59    &   -11.32  $\pm$   2.86    &   -15.87  $\pm$   2.20    \\
RX   J0452.5+1730   &   -   &   -132    &   -6  &   -39 &   -13.30  $\pm$   2.26    &   -10.14  $\pm$   4.63    &   -11.33  $\pm$   3.34    \\
RX   J0457.0+1517   &   -   &   -116    &   -11 &   -35 &   -11.71  $\pm$   1.96    &   -9.60   $\pm$   1.48    &   -10.03  $\pm$   1.17    \\
RX   J0457.5+2014   &   -   &   -131    &   -3  &   -32 &   -12.80  $\pm$   1.98    &   -10.28  $\pm$   1.91    &   -9.10   $\pm$   1.55    \\
\hline
\end{tabular}
}
\end{center}
\end{table*}

\subsubsection{Kinematic properties of group members with known radial velocities}

Table~\ref{VradPar} lists the parallaxes for core group members with
known radial velocities. Columns 1 and 2 list star names together
with their HBC number, while Columns 3 and 4 give the individual
parallax and distance (with their uncertainties) computed from the
individual stellar radial velocity in addition to the proper motion
information. The symbol ``:'' indicates uncertain values for two
stars that will be discussed in Sect.~\ref{DiscussionSection}.

\begin{figure}
\resizebox{\hsize}{!}{\includegraphics[angle=0]{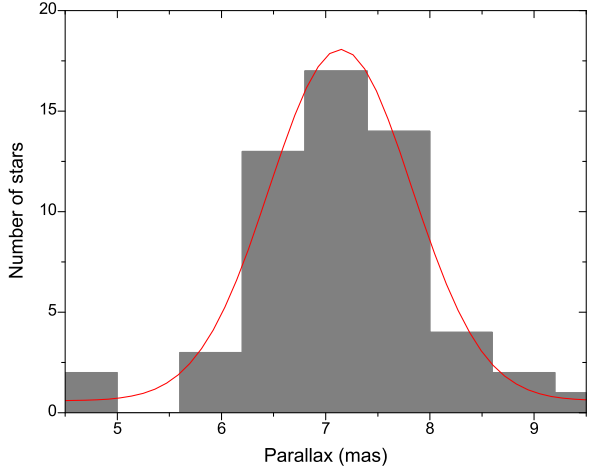}}
\caption[]{Histogram of the parallaxes for group members with known
radial velocities. The superimposed curve represents a Gaussian fit
to the data.} \label{Fig15}
\end{figure}

A histogram with bin size equal to 0.6 mas of derived parallaxes for
stars with known radial velocities is displayed in Fig.~\ref{Fig15}
together with a Gaussian fit to the data. The peak of the Gaussian
curve is at $\pi = 7.14 \pm 0.04$ mas, corresponding to a distance
of 140 pc, and its HWHM is equal to $0.68 \pm 0.05$ mas. The reduced
$\chi^2$ of the Gaussian fit is 1.16. The apparently normal
distribution of the data suggests that no systematic bias affects
the derived parallaxes. The average values and standard deviations
of the Galactic velocity components $U, V, W,$ and $V_{space}$ are

\medskip
\noindent $\left\{ \begin{array}{lll}
U & = &  -16.45 \pm 4.56 \: \rm{km/s}  \\
V & = &  -13.18 \pm 2.52 \: \rm{km/s}  \\
W & = &  -10.97 \pm 4.04 \: \rm{km/s}  \\
V_{space} & = & 23.95 \pm  5.87 \: \rm{km/s}.
\end{array}
\right.$
\medskip

Table~\ref{GroupGalVel} lists the name and HBC number, as well as
the values of the derived Galactic coordinates (Columns 3 to 5) and
Galactic velocity components with their uncertainties (Columns 6 to
8) for the 67 members of the moving group -- both core members and
confirmed candidates (see Sect.\ref{AddCandidates} below) -- with
known radial velocities,  i.e., those stars for which the derived
parallaxes are most accurate. This table can be used to identify
individual group members in the figures.

Figure~\ref{Fig12} displays the projected velocities on the $XY$,
$XZ$, and $YZ$ planes for the 67 same group members.
Figure~\ref{Fig13} shows their spatial distribution. The stars are
distributed in a $X,Y,Z$ cube with sides of approximately 60pc, and
filamentary structure, reminiscent of the Taurus molecular cloud
structure, can be seen in all three $(X,Y)$, $(Y,Z)$, and $(X,Z)$
projection planes. In the $(X,Z)$ plane, there is some tendency for
the densest subgroups of stars to align roughly in the direction
away from the Sun. This is an expected result of the changed
parallax values, since the stars must maintain their apparent
distribution in the plane of the sky, as seen from Earth. Note that
individual stars can be identified by their coordinate values given
in Table~\ref{GroupGalVel}. Finally, Fig.~\ref{Fig14} displays
histograms of the $U, V, W$ galactic velocity components and spatial
velocities of the members of the Taurus moving group with known
radial velocities.

\renewcommand{\arraystretch}{1.2}
\begin{table}
\caption{Tentative parallaxes (in mas) and distances (in pc) of 25
moving group stars with unknown radial velocities computed under the
assumption that they share the same spatial
velocity.}\label{VspacePar}
\begin{center}
\tiny{
\begin{tabular}{lcrr}
\hline\hline Star  &   HBC  &  $\pi_{Vspace} \pm
\sigma_\pi$ & $d_{Vspace} \pm \sigma_d$ \\
\hline
LkCa 1  &   365 &   8.86    $\pm$   2.69    &   113 $^{+    49  }_{ -26 }$\\
FN   Tau    &   24  &   9.64    $\pm$   2.84    &   104 $^{+    43  }_{ -24 }$\\
LkCa 3  &   368 &   5.89    $\pm$   2.16    &   170 $^{+    98  }_{ -46 }$\\
FO   Tau    &   369 &   6.76    $\pm$   2.12    &   148 $^{+    68  }_{ -35 }$\\
V892 Tau    &   373 &   8.70    $\pm$   2.51    &   115 $^{+    47  }_{ -26 }$\\
Haro 6-5B   &   381 &   5.19    $\pm$   2.31    &   193 $^{+    154 }_{ -59 }$\\
FS   Tau    &   383 &   5.10    $\pm$   2.30    &   196 $^{+    161 }_{ -61 }$\\
LkCa 21 &   382 &   8.72    $\pm$   2.52    &   115 $^{+    46  }_{ -26 }$\\
FV Tau/c    &   387 &   8.76    $\pm$   2.91    &   114 $^{+    57  }_{ -28 }$\\
DG   Tau    &   37  &   6.82    $\pm$   1.75    &   147 $^{+    51  }_{ -30 }$\\
ZZ   Tau    &   46  &   8.53    $\pm$   2.71    &   117 $^{+    55  }_{ -28 }$\\
LkH$\alpha$ 358 &   394 &   5.91    $\pm$   2.73    &   169 $^{+    145 }_{ -53 }$\\
XZ   Tau    &   50  &   7.25    $\pm$   1.89    &   138 $^{+    48  }_{ -28 }$\\
Haro 6-13   &   396 &   6.30    $\pm$   2.56    &   159 $^{+    108 }_{ -46 }$\\
FY   Tau    &   401 &   3.52    $\pm$   1.70    &   284 $^{+    263 }_{ -92 }$\\
FZ   Tau    &   402 &   10.51   $\pm$   2.95    &   95  $^{+    37  }_{ -21 }$\\
UZ   Tau AB &   52  &   7.43    $\pm$   2.49    &   135 $^{+    68  }_{ -34 }$\\
V807 Tau    &   404 &   6.57    $\pm$   1.70    &   152 $^{+    53  }_{ -31 }$\\
IS   Tau    &   59  &   8.34    $\pm$   2.65    &   120 $^{+    56  }_{ -29 }$\\
HO   Tau    &   64  &   5.84    $\pm$   2.31    &   171 $^{+    112 }_{ -49 }$\\
FF   Tau    &   409 &   5.80    $\pm$   2.30    &   172 $^{+    113 }_{ -49 }$\\
CoKu Tau-Aur Star 3 &   411 &   9.16    $\pm$   3.04    &   109 $^{+    54  }_{ -27 }$\\
CoKu HP Tau G2  &   415 &   5.38    $\pm$   1.44    &   186 $^{+    68  }_{ -39 }$\\
CoKu LkH$\alpha$ 332 G1 &   423 &   8.73    $\pm$   2.42    &   115 $^{+    44  }_{ -25 }$\\
DP   Tau    &   70  &   6.54    $\pm$   2.36    &   153 $^{+    87  }_{ -41 }$\\
\hline
\end{tabular}
}
\end{center}
\end{table}

\subsubsection{Approximate parallaxes for other moving group members}

We now assume that all members of the moving group share the same
average value of $V_{space}$ derived above. This hypothesis allows
us to compute \emph{expected} radial velocities for all stars in the
moving group and subsequently to derive tentative parallaxes for
those group members.

\begin{figure}
\resizebox{\hsize}{!}{\includegraphics[angle=0]{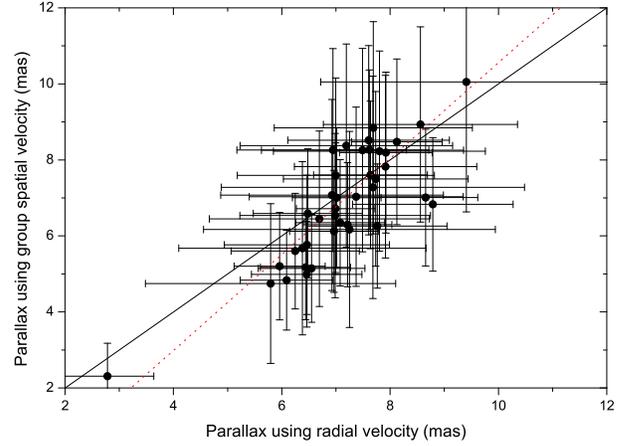}}
\caption[]{Comparison of parallaxes computed using the radial
velocity data and the group spatial velocity derived from the
subgroup of stars with known radial velocities. The solid line
indicates perfect correlation, while the dotted line represents a
linear fit to the data.} \label{Fig16}
\end{figure}

Figure~\ref{Fig16} compares the parallax values obtained by both
methods and shows that they give similar results within the error
bars, but the scatter is large. There is also a bias in the derived
parallaxes in the sense that small parallaxes are underestimated,
and large parallaxes are overestimated, when determined from the
group spatial velocity, as shown by the linear fit to the data
(dotted line in Fig.~\ref{Fig16}). According to this regression
curve, the parallax average and standard deviation computed using
radial velocities, $6.99 \pm 1.14$ mas, translate to $6.74 \pm 1.44$
mas when using the average spatial velocity for computing the
parallax; not only is the resulting parallax smaller on average but
its standard deviation is larger. The parallaxes deviating most from
the mean in Table~\ref{VspacePar} should therefore be viewed with
extreme caution. A case in point is the couple FY/FZ~Tau, two stars
separated by less than 20\arcsec\ on the sky whose derived
parallaxes are respectively the smallest and the largest of our
sample. The result for FY~Tau is obviously unrealistic, as both
stars are seen in projection on the B18 nebula and have the same
line-of-sight visual extinction.

The average parallax and associated standard deviation of stars in
Table~\ref{VspacePar} are $\pi = 7.21 \pm 1.71$ mas, corresponding
to an average distance of 139 pc. The average distance of these
stars is similar to that of stars for which we know the radial
velocities, but the parallax standard deviation is almost twice as
large, as anticipated from the regression analysis. Although the
derived parallaxes can only be considered as tentative and would be
much improved, and their uncertainties reduced, if accurate
spectroscopic radial velocities were available for the full sample,
this approximation nevertheless may provide a useful first estimate
of the distance to those moving group members whose parallax is not
too different from the mean (e.g., DG~Tau, XZ~Tau, and UZ~Tau).

\subsection{Additional group member candidates} \label{AddCandidates}

After finding a core moving group among the pre-main sequences stars
of Taurus-Auriga, we now look for plausible additional members among
stars in the \cite{2005A&A...438..769D} catalog whose evolutionary
status is uncertain. We thus now consider the full sample of 217
stars discussed in Sect.~\ref{CatalogDescriptionSection} and
identify those whose space motion is compatible with that of the
core moving group.

\subsubsection{Analysis and results}

Specifically, we assumed that the CP of the entire moving group has
the coordinates found above for the core moving group and computed
the individual probability of each star being part of the moving
group converging to this point by using Eq.~\ref{indivproba}. A
$p_{min}$ minimum value of 0.91 was chosen because it is the lowest
value to allow for the rejection of all those stars in the subgroup
of confirmed PMS objects that are not actual members of the core
moving group. Stars with a lower $p_{min}$ were thus eliminated,
leaving us with 30 stars whose space motion is compatible with that
of the core moving group.

Among these 30 objects, 15 have known radial velocities; their
parallaxes are given in Table~\ref{CandVradPar} and the kinematic
properties of confirmed members (see below) are listed in
Table~\ref{GroupGalVel}. As previously, we computed the spatial
velocity of this sample to compute approximate parallaxes for the
remaining 15 stars, and we give these tentative parallaxes in
Table~\ref{CandVspacePar}.

\renewcommand{\arraystretch}{1.2}
\begin{table}
\caption{Parallaxes (in mas), distances (in pc), and probable
evolutionary status of 15 possibly additional moving group stars
with known radial velocities. Parallaxes of stars that are not PMS
members of Taurus-Auriga are meaningless and are therefore not given
here.}\label{CandVradPar}
\begin{center}
\tiny{
\begin{tabular}{lrrc}
\hline\hline Star  &  $\pi_{Vspace} \pm
\sigma_\pi$ & $d_{Vspace} \pm \sigma_d$ & PMS\\
\hline
RX   J0400.5+1935   &       &     &   n\\
GSC  01262-00421    &   6.03    $\pm$   0.88    &   166 $^{+    28  }_{ -21 }$  &   y\\
RX   J0405.7+2248   &   5.69    $\pm$   0.87    &   176 $^{+    32  }_{ -23 }$  &   y\\
RX   J0406.7+2018   &   6.30    $\pm$   0.95    &   159 $^{+    28  }_{ -21 }$  &   y\\
RX   J0406.8+2541   &       &    &   n\\
RX   J0423.7+1537   &   7.56    $\pm$   1.08    &   132 $^{+    22  }_{ -16 }$  &   y\\
RX   J0424.8+2643B  &      &     &   n\\
RX   J0432.8+1735   &   6.47    $\pm$   2.37    &   155 $^{+    89  }_{ -41 }$  &   y\\
V1078 Tau  &   3.87    $\pm$   0.71    &   259 $^{+    58  }_{ -40 }$  &   y\\
RX   J0444.9+2717   &      &     &   n\\
RX   J0450.0+2230   &      &     &   n\\
RX   J0452.5+1730   &   7.24    $\pm$   2.62    &   138 $^{+    78  }_{ -37 }$  &   y\\
RX   J0455.8+1742   &       &     &   n\\
RX   J0457.0+1517   &   8.22    $\pm$   1.13    &   122 $^{+    19  }_{ -15 }$  &   y\\
RX   J0457.5+2014   &   7.44    $\pm$   1.06    &   134 $^{+    22  }_{ -17 }$  &   y\\
\hline
\end{tabular}
}
\end{center}
\end{table}
\renewcommand{\arraystretch}{1.2}
\begin{table}
\caption{Tentative parallaxes (in mas), distances (in pc), and
probable or tentative evolutionary status of 15 possibly additional
moving group stars with unknown radial velocities. Parallaxes of
stars that are not PMS members of Taurus-Auriga are meaningless and
are therefore not given here. Confirmation of PMS status is required
for most objects.}\label{CandVspacePar}
\begin{center}
\tiny{
\begin{tabular}{lrrc}
\hline\hline Star  &  $\pi_{Vspace} \pm
\sigma_\pi$ & $d_{Vspace} \pm \sigma_d$ & PMS\\
\hline
1RXS J035330.5+263152   &      &   &   n\\
RX   J0412.8+2442    &   &   & n\\
KPNO-Tau 11 &   8.15    $\pm$   2.75    &   123 $^{+    63  }_{ -31 }$  &   y\\
FR   Tau    &   7.25    $\pm$   2.58 (?)    &   138 $^{+    76  }_{ -36 }$ (?)  &   y?\\
FU   Tau    &   8.44    $\pm$   2.87 (?)    &   118 $^{+    61  }_{ -30 }$ (?) &   y?\\
FW   Tau    &   6.81    $\pm$   2.58 (?)    &   147 $^{+    89  }_{ -40 }$ (?)  &   y?\\
RX   J0431.3+2150   &   4.39    $\pm$   1.23    &   228 $^{+    89  }_{ -50 }$  &   y\\
EZ   Tau    &     &    &   n\\
FI   Tau    &   7.61    $\pm$   2.79 (?)    &   131 $^{+    76  }_{ -35 }$ (?)  &   y?\\
GN   Tau    &   6.07    $\pm$   2.50 (?)    &   165 $^{+    115 }_{ -48 }$ (?)  &   y?\\
Haro 6-36   &   5.83    $\pm$   2.59 (?)   &   171 $^{+    137 }_{ -53 }$ (?)  &   y?\\
Haro 6-39   &   7.92    $\pm$   2.88 (?)    &   126 $^{+    72  }_{ -34 }$ (?)  &   y?\\
RX   J0456.6+3150   &   2.68    $\pm$   0.84 (?)    &   373 $^{+    170 }_{ -89 }$ (?)  &   ?\\
1RXS J050029.8+172400   &   4.92    $\pm$   1.39 (?)    &   203 $^{+    80  }_{ -45 }$ (?)  &   y?\\
1RXS J051111.1+281353   &   6.82    $\pm$   1.75 (?)   &   147 $^{+    51  }_{ -30 }$ (?)  &   ?\\
\hline
\end{tabular}
}
\end{center}
\end{table}

\subsubsection{Evolutionary status of candidate members}

The ROSAT-detected stars included in Table~\ref{CandVradPar} were
studied in some detail by \cite{2000A&A...359..181W}, who assessed
their evolutionary status on the basis of their lithium line
strengths. The two other stars, GSC 01262-00421 and V1078 Tau, were
studied by \cite{1988AJ.....96..297W}, who concluded, also from
their lithium strength, that they were \emph{bona fide} PMS objects.
A ``y'' in last column of Table~\ref{CandVradPar} indicates that the
star is a PMS object on the basis of these two works. We thus find
that 9 of of the 15 stars are confirmed PMS objects and thus members
of the Taurus-Auriga moving group.

Most stars included in Table~\ref{CandVspacePar} have not been
studied in any kind of detail, so assessing their evolutionary
status on the basis of the available information is a challenge. All
stars marked with ``y?" in the last table column are IRAS point
sources \citep{1989AJ.....97.1451S} and are therefore likely to be
associated with circumstellar matter, which strongly hints at PMS
status; however, spectroscopic confirmation is unavailable.
1RXS~J035330.5+263152 is the X-ray counterpart of a likely member of
the Pleiades corona \citep{1975A&A....43..423P}, while RX
J0412.8+2442 is unlikely to be young according to
\cite{1997A&AS..124..449M}. KPNO-Tau~11 is a recently-detected
low-mass member of the Taurus star-forming region
\citep{2003ApJ...590..348L}, while RX~J0431.3+2150 was originally
proposed as a PMS object by \cite{1996A&A...312..439W}, and this
status was confirmed by \cite{1997A&A...318..495B}. EZ~Tau is a
flare star and probable low-mass member of the Hyades
\citep{1993MNRAS.265..785R}. 1RXS~J050029.8+172400 was classified as
a T~Tauri star on the basis of its X-ray variability by
\cite{2003A&A...403..247F}, but this needs to be confirmed by
optical spectroscopy. Finally, there is not enough information to
assess the status of RX~J0456.6+3150 and 1RXS~J051111.1+281353. The
second star was classified as T~Tauri star by
\cite{2004ChJAA...4..258L} on the basis of its proper motion, but we
have seen that this criterion is not discriminatory. Altogether,
only 2 of these 15 stars are confirmed members of the moving group,
while 8 are possibly additional members.

\subsection{Note on X-ray selected stars located south of
Taurus}

Finally, we searched for putative moving group members in the region
south of Taurus where a large population of X-ray sources possibly
related to the Taurus-Auriga star-forming region was detected by the
ROSAT All-Sky X-ray survey \citep{1997A&AS..124..449M}. We thus
considered the region $3^h 50^m$ \simlt $\alpha(2000)$ \simlt $5^h
10^m$ and $0$\deg \simlt $\delta(2000)$ \simlt $15$\deg, where the
\cite{2005A&A...438..769D} catalog lists 47 stars, and did the same
analysis as in Sect.~\ref{AddCandidates} thus searching for
additional members of the core moving group defined in
Sect.~\ref{CoreGroup}. In this way, we identified 7 stars whose
space motion is compatible with that of the core moving group. These
are RX~J0357.3+1258, RX~J0358.1+0932,RX~J0404.4+0519,
RX~J0450.0+0151, RX~J0405.5+0324, RX~J0441.9+0537, and RX
J0445.3+0914.

Since the kinematic analysis is not sufficient for identifying
moving group members with any kind of certainty, we searched the
literature to determine the evolutionary status of these objects.
The first four objects were studied among other PMS candidates by
\cite{1997A&A...325..647N}, who found that RX~J0450.0+0151 is a
likely PMS star while the other stars are on the main sequence. The
remaining stars are part of a sample studied by
\cite{1997A&AS..124..449M}, who list them as main sequence objects.

With the possible exception of RX~J0450.0+0151, for which we derive
a parallax of $11.84 \pm 1.33$ mas corresponding to a distance of
$84^{+11}_{-9}$ pc, we thus conclude that the X-ray sources located
south of Taurus that have proper motions compatible with those of
the core Taurus-Auriga moving group are field stars unrelated to the
Taurus-Auriga PMS association.

\subsection{A posteriori assessment of results}

To assess the validity of the results presented above, we performed
additional Monte Carlo simulations of the core moving group search
using the results derived above. For this, we constructed synthetic
data sets as explained in Sect.~\ref{MCSims}; but instead of drawing
a random velocity vector, we used the velocity of the moving group
as derived from the CP search, i.e., $V_{group} = 7.3$ km/s, $\Theta
= 116$\deg, and $\Phi = 171$\deg. To define the individual velocity
of each synthetic star, we added to the group velocity vector an
individual random velocity drawn from a normal distribution with
variance $\sigma_{int}$ and derived the resulting proper motions and
radial velocities from there. We then added random measurements
errors and corrected for the Galactic rotation as explained in
Sect.~\ref{MCSims}. Because the considered sample of 117
Taurus-Auriga stars used in the CP search is made up of confirmed
pre-main sequence stars\footnote{Except for Wa Tau/1 (HBC 408),
whose pre-main sequence status is questionable but which was duly
eliminated in the CP search.}, we did not include interlopers in
these simulations.

\begin{figure}
\resizebox{\hsize}{!}{\includegraphics[angle=0]{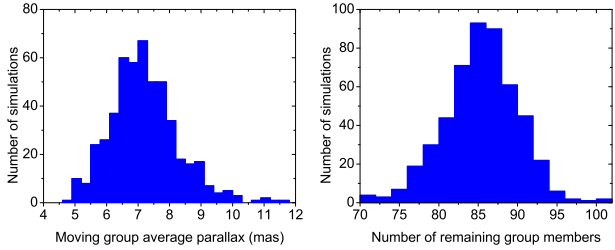}}
\caption[]{Results of simulations performed with the computational
parameters used in Sect.~\ref{KinematicAnalysisSection}. Histograms
of the recovered average parallaxes and of the numbers of recovered
group members are shown.} \label{Fig17}
\end{figure}

\begin{figure}
\resizebox{\hsize}{!}{\includegraphics[angle=0]{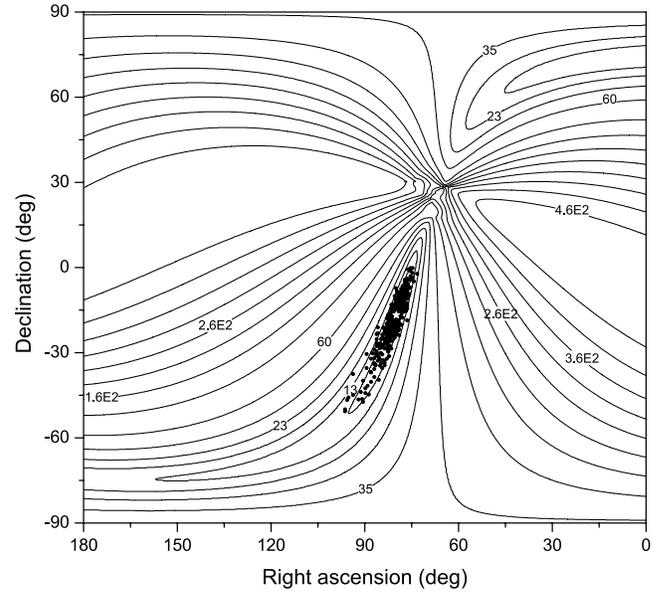}}
\caption[]{Realizations of CP coordinates (dots) in the Monte-Carlo
simulation,  overlaid on the $X^2$ contours for the core moving
group solution derived in Sect.~\ref{KinematicAnalysisSection}. The
white star denotes the CP coordinates of the Taurus-Auriga core
moving group as derived in that section.} \label{Fig18}
\end{figure}

We constructed 500 such realizations in our simulation, which we ran
with the same computational parameters as the actual CP search of
Sect.~\ref{KinematicAnalysisSection}. The results of these
simulations are illustrated by Fig.~\ref{Fig17}, which demonstrates
that the probability of recovering the average parallax of a moving
group with Taurus-Auriga properties is 0.38 and that the average
number of stars in the recovered moving group is $85 \pm 5$, whereas
we found 83 members for the core moving group in the actual
solution. The probability that the CP method cannot find a moving
group in this situation is 0.05. The histogram of derived parallaxes
can be fitted by a Gaussian curve peaking at $7.06 \pm 0.04$ mas
with HWHM equal to $0.86 \pm 0.06$ mas, whereas the input values for
these two quantities were respectively 7.31 and 0.49 mas. Our actual
computation recovers the average parallax of Taurus because we hit
on one of the best possible solutions for the CP coordinates, as
shown below.

The derived CP coordinates in the 500 simulations are shown in
Fig.~\ref{Fig18}. As previously discussed, e.g., by
\cite{1999MNRAS.306..381D} CP coordinates fall along the great
circle associated with the CP, which is accurately defined by the
proper motions, but the precise location of the CP on this great
circle is quite uncertain. For comparison, we also plot in
Fig.~\ref{Fig18} the $X^2$ values (as defined by Eq.~\ref{X2}) that
measure the probability of finding the CP at a given location on the
plane of the sky. The curves plotted here correspond to the solution
found in Sect.~\ref{KinematicAnalysisSection}, and the white star
indicates the coordinates of the derived Taurus-Auriga core moving
group CP. The topology of the $X^2$ surface seen here is typical of
the CP method. A large fraction of the sky forms a high plateau (in
terms of $X^2$) cut by a deep valley following the great circle
associated with the CP, with an elongated minimum following the
great circle at the location of the CP. The height of the plateau
and the depth of the valley are directly related to the velocity
dispersion and measurements uncertainties in the sense that reducing
these values increases the contrast between plateau and valley, thus
making it easier to locate the CP precisely.

The lowest $X^2$ contour in Fig.~\ref{Fig18} corresponds to a value
of 13 and includes most CP realizations. With an $X^2$ of 10.6, the
coordinates of the CP derived for Taurus-Auriga obviously represent
one of the best possible solutions, thus validating the choice of
computational parameters used to find it. However, we expect only
73\% of the actual moving group members to be recovered in the
computation. Therefore, most stars in our core sample are likely to
be actual members of the moving group even though they were rejected
during the CP search. In order to identify these potentially
additional moving group members and derive their individual
parallaxes, more precise radial velocities and proper motions will
be necessary.

\section{Remarks on positions and parallaxes}\label{DiscussionSection}

A first check of our results is provided by a comparison with
Hipparcos parallaxes.

\subsection{Comparison with Hipparcos results}

As already mentioned in Sect.~\ref{CatalogDescriptionSection}, there
is little overlap between Taurus YSOs and Hipparcos targets, due
mainly to the faintness of most Taurus PMS stars. We found 11 stars
among our moving group that have been observed by Hipparcos and
listed them in Table~\ref{HIPComp}, together with the parallaxes
derived in this work and the Hipparcos parallaxes. The values agree
within the error bars except for 3 objects: BP~Tau, DF~Tau, and
UX~Tau, for which Hipparcos reports a negative parallax. The
discrepant Hipparcos parallaxes of BP~Tau and DF~Tau were discussed
in some detail by \cite{1999A&A...352..574B}, who concluded from a
re-analysis of the Hipparcos data that the derived large parallaxes
were unlikely to be significant, although they could not be ruled
out entirely. \cite{1999A&A...352..574B} also re-computed a more
precise parallax of $7.98 \pm 3.15$ mas for RW~Aur after rejecting
the bad abscissae. We thus conclude that there is a general
agreement between the parallaxes derived here and the Hipparcos
values except for BP~Tau and DF~Tau. The values reported in this
work for these two objects are much more in line with expectations
than the Hipparcos values, since the two stars are clearly
associated with the molecular cloud.
\renewcommand{\arraystretch}{1.0}
\begin{table}
\caption{Comparison with Hipparcos catalog parallaxes. Units are
mas.}\label{HIPComp}
\begin{center}
\tiny{
\begin{tabular}{lccrr}
\hline\hline Name        &   HIP &   HBC &   $\pi_{Vrad} \pm \sigma_\pi$            &   $\pi_{HIP} \pm \sigma_\pi$         \\
\hline
V773    Tau &   19762   &   367 &   6.84    $\pm$   1.07    &   9.88    $\pm$   2.71    \\
V410    Tau &   20097   &   29  &   7.3 $\pm$   0.88    &   7.31    $\pm$   2.07    \\
BP  Tau &   20160   &   32  &   7.46    $\pm$   0.66    &   18.98   $\pm$   4.65    \\
RY  Tau &   20387   &   34  &   7.62    $\pm$   0.68    &   7.49    $\pm$   2.18    \\
HD  283572  &   20388   &   380 &   7.64    $\pm$   1.05    &   7.81    $\pm$   1.3 \\
DF  Tau &   20777   &   36  &   7.76    $\pm$   1.29    &   25.72   $\pm$   6.36    \\
UX  Tau &   20990   &   43  &   6.55    $\pm$   0.99    &   -6.68   $\pm$   4.04    \\
AB  Aur &   22910   &   78  &   8.79    $\pm$   1.48    &   6.93    $\pm$   0.95    \\
SU  Aur &   22925   &   79  &   6.99    $\pm$   0.73    &   6.58    $\pm$   1.92    \\
RW  Aur &   23873   &   80  &   7.21    $\pm$   0.82    &   14.18   $\pm$   6.84    \\
RX  J0406.7+2018    &   19176   &   -   &   6.3 $\pm$   0.95    &   6.43    $\pm$   1.84    \\
\hline
\end{tabular}
}
\end{center}
\end{table}

\begin{figure}
\resizebox{\hsize}{!}{\includegraphics[angle=0]{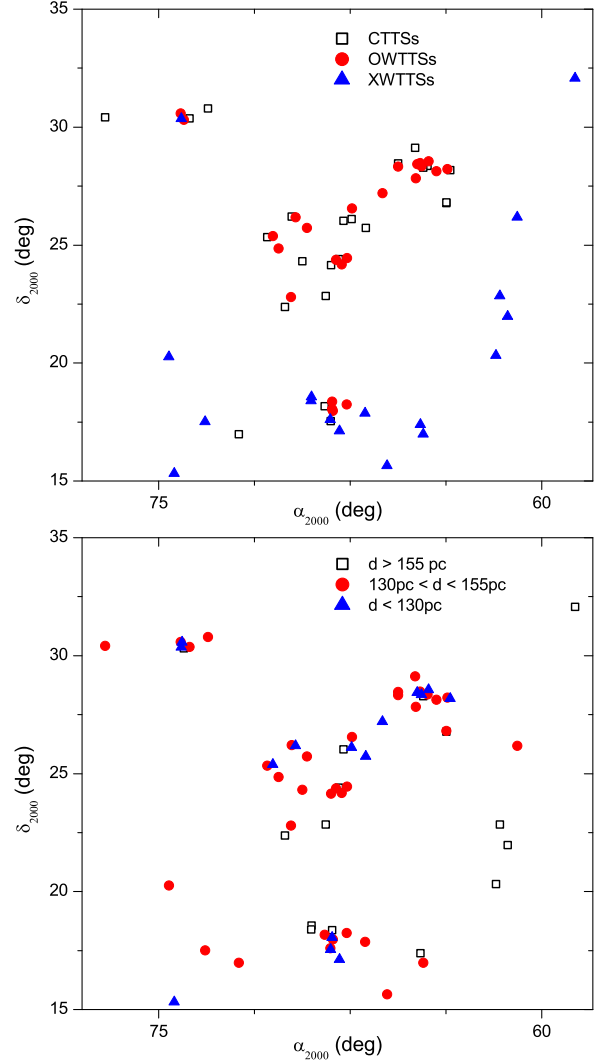}}
\caption[]{Location of the moving group stars in the plane of the
sky. As indicated in the legend, different symbols mark the stars
closer than 130pc, those farther away than 155pc, and the remaining
stars.} \label{Fig19}
\end{figure}

We restrict the following discussion to those members of the moving
group with measured radial velocities since we derived accurate
parallaxes for these objects, whereas the parallaxes derived for
other stars from the group's average spatial velocity are tentative.

\subsection{Notes on remarkable stars}

NTTS 035120+3154 (HBC 352/353) provides an interesting illustration
of the current limitations of the study reported here. According to
\cite{1993A&A...278..129L}, the SW and NE components of this object
are likely to form a physical pair. The proper motions and radial
velocities of both stars, given in Table~\ref{NTTS}, agree within
the error bars; and yet, we derive distances (see
Table~\ref{VradPar}) that, while they agree with each other within
the derived error bars, differ by 34 pc, i.e., 14\% of the mean
distance value of 242 pc. This uncertainty thus appears
representative of what can be done with the current data. Assuming
that the radial velocities were equal within 0.1 km/s, one finds
that the gap between distances is reduced to 22 pc, or 9\% of the
mean distance. This increased accuracy is thus within easy reach
since it only requires accurate radial velocities. On the other
hand, this also shows that without much more accurate proper motions
measurements, the uncertainty on derived distances cannot be
expected to decrease much below 10\%.

\begin{table}
\caption{Proper motions (in mas/yr) and radial velocity (in km/s)
for NTTS 035120+3154}\label{NTTS}
\begin{center}
\tiny{
\begin{tabular}{lccc}
\hline\hline Name & $\mu_\alpha \cos \delta \pm \sigma_\mu$ &
$\mu_\delta \pm \sigma_\mu  $ & $V_{rad} \pm \sigma_V $\\
\hline
NTTS 035120+3154SW & 8 $\pm$  2  & -9 $\pm$ 2  & 16.0   $\pm$ 1.5 \\
NTTS 035120+3154NE & 7 $\pm$  2  & -11 $\pm$ 2  & 15.1 $\pm$ 1.5 \\
\hline
\end{tabular}
}
\end{center}
\end{table}

NTTS 042835+1700 (HBC 392) appears to be the closest star in the
moving group at 106 pc (2.1$\sigma$ from the average moving group
distance), while NTTS 043124+1824 (HBC 407) is the farthest away at
360 pc, i.e., 3.7$\sigma$ away from the average. The proper motion
values of HBC 407, an apparently single star
\citep{2001A&A...369..971K}, are indeed very small, with $\mu_\alpha
\cos \delta = 0 \pm 2$ mas/yr and $\mu_\delta = -7 \pm 2$ mas/yr.
Its radial velocity of 18.4 km/s is not very different from the
average of 16 km/s and it is therefore difficult to understand how
the star could have reached this location in space if it is a true
member of the Taurus-Auriga moving group. Its PMS nature has been
confirmed by \cite{1988AJ.....96..297W} and others, and it is thus
unlikely to be an interloper. Its discrepant parallax, if true,
remains therefore somewhat of a mystery, but we note that the
computed uncertainty on its value is particularly high.

CZ~Tau (HBC 31) is another interesting case because it is located at
less than 20\arcsec\  from DD~Tau and has the same proper motion yet
its radial velocity is highly discrepant (compared to the
Taurus-Auriga average) at 44 km/s. The radial velocity measurements
for both objects go back to the classic paper where
\cite{1949ApJ...110..424J} discussed the class of T~Tauri stars for
the first time, and they are noted as very uncertain in the
\cite{1988cels.book.....H} catalog. The derived parallaxes are
therefore also very uncertain so we marked them as such in
Table~\ref{VradPar}.

\subsection{Properties of parallaxes for various YSO sub-classes}

\begin{figure}
\resizebox{\hsize}{!}{\includegraphics[angle=0]{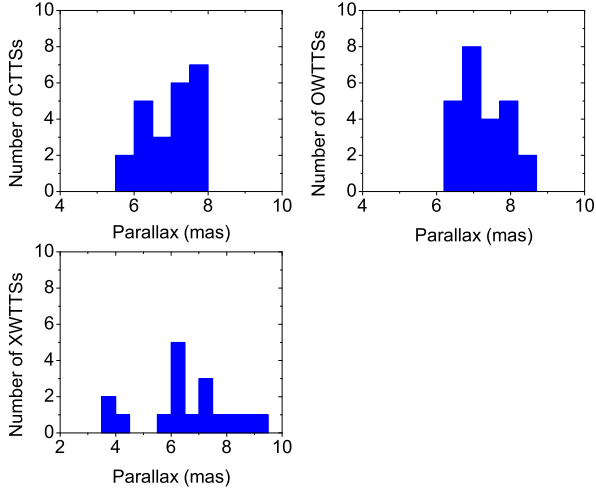}}
\caption[]{Histograms of parallaxes computed using the radial
velocities for the sub-samples of CTTSs, OWTTSs, and XWTTSs in the
moving group.} \label{Fig20}
\end{figure}

Figure~\ref{Fig19} displays the members of the moving group as seen
in the plane of the sky. The moving group contains 23 classical
T~Tauri stars (CTTSs) and 43 confirmed weak emission-line T~Tauri
stars (WTTSs) with known radial velocities. Among the WTTSs, we have
18 X-ray selected WTTSs (XWTTSs) and 25 optically-selected WTTSs
(OWTTSs). A comparison of the parallaxes of stars in these subgroups
confirms that X-ray selected WTTSs tend to be found on the outskirts
of the molecular clouds where extinction is relatively low, while
CTTSs (and OWTTSs) are associated with the denser regions of the
clouds (see also the parallax histograms of the various YSO
populations in Fig.~\ref{Fig20}).

An interesting, although not unexpected new finding of our
investigation is that WTTSs are located not only in front of the
clouds, but also at their back, while CTTSs are confined to the
central parts of the moving group. This is illustrated by
Fig.~\ref{Fig20}, which displays histograms of the CTTS and WTTS
parallaxes. All CTTSs are at distances between 126 and 173 pc, while
WTTSs\footnote{We excluded here the two stars discussed above, CZ
Tau and NTTS 043124+1824, because their parallaxes are highly
uncertain.} span the range of distances between 106 and 259 pc.

Computing average parallaxes and standard deviations for these
various subgroups, we get
\begin{eqnarray}
\overline{\pi_{CTTS}} & = &  7.01 \pm 0.67 {\rm mas} \nonumber \\
\overline{\pi_{WTTS}} & = &  7.00 \pm 1.15 {\rm mas}  \nonumber \\
\overline{\pi_{OWTTS}} & = &  7.28 \pm 0.65 {\rm mas} \nonumber \\
\overline{\pi_{XWTTS}} & = &  6.60 \pm 1.55 {\rm mas}, \nonumber
\end{eqnarray}
where we note that the standard deviation of the XWTTS parallaxes is
more than twice as large as those of CTTSs and OWTTSs.

For the sample of CTTSs and OWTTSs with known radial velocities, the
individual parallaxes of which are presumably more accurate than
those computed from the group spatial velocity, we recover the
average parallax value derived by \cite{1999A&A...352..574B} from
Hipparcos data. Our analysis thus confirms that the core of the
Taurus association, as defined by this CTTS and OWTTS sample, is
located at $140^{+14}_{-12}$ pc.

\begin{figure}
\resizebox{\hsize}{!}{\includegraphics[angle=0]{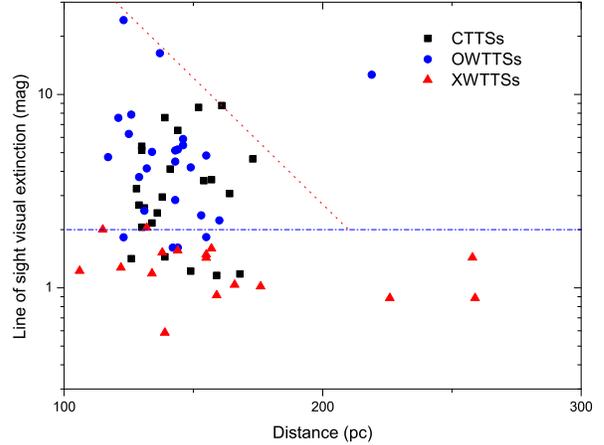}}
\caption[]{Line-of-sight visual extinction of moving group members
plotted as a function of their distance. Different symbols, shown in
the legend, are used to identify X-ray selected WTTSs (XWTTSs),
optically selected WTTSs (OWTTSs), and CTTSs. The meaning of the two
lines is discussed in the text.} \label{Fig21}
\end{figure}

That we are actually observing WTTSs on both sides of the molecular
clouds may seem at first to contradict the idea that X-ray selected
stars tend to be detected in low-extinction regions. However, one
must recall that extinction is very patchy in the Taurus-Auriga
star-forming region, characterized by a dense filamentary structure
and zones of lower molecular gas density. To test whether XWTTSs are
indeed preferentially found along low-extinction lines of sight, we
used the extinction maps of \cite{1998ApJ...500..525S} to derive the
line-of-sight (LOS) visual extinction in the direction of each
moving group star. Figure~\ref{Fig21} shows that XWTTSs are on the
average located on lines of sight that have lower visual extinctions
than OWTTSs and CTTSs. The horizontal dash-dotted line indicates the
upper limit of LOS extinction for XWTTSs of 2 mag. Not surprisingly,
there is no significant difference between the LOS extinctions of
OWTTS and CTTS subgroups. Note that no star farther away than
$\approx 180$ pc (except for one star, CZ~Tau, the parallax of which
is highly uncertain as discussed above) has a LOS extinction higher
than 1.5 magnitudes. The upper envelope of the data points in
Fig.~\ref{Fig21} (dotted line) indicates approximately how far one
``sees'' in the Taurus star-forming region for a given LOS
extinction. Lines of sight where the extinction is in the range $2
\leq A_V \leq 20$ mag. become opaque in the optical domain at
distances in the range $210 \geq d \geq 120$ pc. From these numbers,
and using the conversion factor from extinction to hydrogen column
density  given by \cite{1995A&A...293..889P}, one finds an average
hydrogen density of $\approx 2 \cdot 10^2$ cm$^{-3}$ in Taurus, a
value that agrees with expectations for large molecular clouds.

\section{Conclusions}\label{ConclusionSection}

We have identified a moving group of 94 stars in the Taurus-Auriga
association that defines the kinematic properties of the
T~association. Because the variant of the CP search method that we
developed for dealing with the high internal velocity dispersion
among Taurus subgroups possibly eliminates a number of potential
group members, we detected a \emph{minimum} moving group that may
not contain all the kinematically associated stars in the region.

Determination of accurate parallaxes for all moving group members is
hampered by the lack of observed high-precision radial velocities
for many group members. New radial velocity measurements for all
stars of the \cite{2005A&A...438..769D} proper-motion catalog are
obviously needed in order to make further progress in deriving
accurate parallaxes for individual YSOs in T~associations. We
therefore encourage observers to make use of the highly efficient
and precise spectrographs employed, in particular, for extra-solar
planet searches to perform a complete radial velocity survey of
Taurus-Auriga and other star-forming regions.

The present result nevertheless represents a first step towards
better understanding the distances to Taurus-Auriga PMS stars, and
we use these new distances to re-assess the physical properties of
members of the Taurus-Auriga moving group in a companion paper
(Bertout \& Siess, in preparation).

\begin{acknowledgements}
We are grateful to Ulrich Bastian, George Herbig, and Steven Shore
for their careful reading of a previous version of the manuscript
and for useful comments. We are indebted to an anonymous referee for
pointing out that the interloper problem was likely to be much more
severe than we had anticipated, which led to a much improved
analysis. This research made advanced use of the Centre de Donn\'ees
de Strasbourg facilities in the framework of a test program aiming
at improving the ease of use and inter-operability of data mining
tools. We acknowledge use of the NASA/ IPAC Infrared Science
Archive, which is operated by the Jet Propulsion Laboratory,
California Institute of Technology, under contract with the National
Aeronautics and Space Administration.
\end{acknowledgements}

\bibliographystyle{aa}
\bibliography{5842}
\end{document}